\documentclass{aa} 
\usepackage{graphicx}
\usepackage{color}
\usepackage{float}
\usepackage{placeins}
\usepackage{natbib}
\bibpunct{(}{)}{,}{a}{}{;}
\usepackage{txfonts}

\begin{document}

%   \title{The ALMA-PILS survey: The -NH$_{2}$ containing molecules hydroxylamine and methylamine and their potential precursor}
   
      \title{The ALMA-PILS survey: Stringent limits on small amines and nitrogen-oxides towards IRAS 16293--2422B}
   \titlerunning{Small amines and nitrogen-oxides towards IRAS 16293--2422B}

   \author{N. F. W. Ligterink\inst{1,2} \and H. Calcutt\inst{3} \and  A. Coutens\inst{4} \and L. E. Kristensen\inst{3} \and T. L. Bourke\inst{5} \and M. N. Drozdovskaya\inst{6} \and H. S. P. M\"{u}ller\inst{7} \and S. F. Wampfler\inst{6} \and M. H. D. van der Wiel\inst{8} \and E. F. van Dishoeck\inst{2,9} \and J. K. J{\o}rgensen\inst{3}}
   \institute { \inst{1}Raymond and Beverly Sackler Laboratory for Astrophysics, Leiden Observatory, Leiden University, PO Box 9513, 2300 RA Leiden,\\ \phantom{'}The Netherlands, E-mail: ligterink@strw.leidenuniv.nl\\      
\inst{2}Leiden Observatory, Leiden University, PO Box 9513, 2300 RA Leiden, The Netherlands\\
  \inst{3}Centre for Star and Planet Formation, Niels Bohr Institute \& Natural History Museum of Denmark, University of Copenhagen,\\ \phantom{'}{\O}ster Voldgade 5--7, 1350 Copenhagen K., Denmark\\
\inst{4}Laboratoire d'Astrophysique de Bordeaux, Univ. Bordeaux, CNRS, B18N, all\'{e}e Geoffroy Saint-Hilaire, 33615 Pessac, France\\
\inst{5}SKA Organization, Jodrell Bank Observatory, Lower Withington, Macclesfield, Cheshire SK11 9DL, UK\\
\inst{6}Center for Space and Habitability (CSH), University of Bern, Sidlerstrasse 5, 3012 Bern, Switzerland\\
\inst{7}I. Physikalisches Institut, Universit\"{a}t zu K\"{o}ln, Z\"{u}lpicher Str. 77, 50937 K\"{o}ln, Germany\\
\inst{8}ASTRON, the Netherlands Institute for Radio Astronomy, Postbus 2, 7990 AA Dwingeloo, The Netherlands\\
\inst{9}Max-Planck Institut f\"{u}r Extraterrestrische Physik (MPE), Giessenbachstr. 1, 85748 Garching, Germany\\
}

   \date{Received}
   % \abstract{}{}{}{}{} 
% 5 {} token are mandatory

\abstract
{Hydroxylamine (NH$_{2}$OH) and methylamine (CH$_{3}$NH$_{2}$) have both been suggested as precursors to the formation of amino acids and are therefore of interest to prebiotic chemistry. Their presence in interstellar space and formation mechanisms, however, are not well established.}
{We aim to detect both amines and their potential precursor molecules NO, N$_{2}$O and CH$_{2}$NH towards the low-mass protostellar binary IRAS 16293--2422, in order to investigate their presence and constrain their interstellar formation mechanisms around a young Sun-like protostar.}
{ALMA observations from the unbiased, high angular resolution and sensitivity Protostellar Interferometric Line Survey (PILS) are used. Spectral transitions of the molecules under investigation are searched for with the CASSIS line analysis software.}
{CH$_2$NH and N$_{2}$O are detected for the first time towards a low-mass source, the latter molecule through confirmation with the single-dish TIMASSS survey. NO is also detected. CH$_{3}$NH$_{2}$ and NH$_{2}$OH are not detected and stringent upper limit column densities are determined.}
{The non-detection of CH$_{3}$NH$_{2}$ and NH$_{2}$OH limits the importance of formation routes to amino acids involving these species. The detection of CH$_{2}$NH makes amino acid formation routes starting from this molecule plausible. The low abundances of CH$_2$NH and CH$_{3}$NH$_{2}$ compared to Sgr B2 indicate that different physical conditions influence their formation in low- and high-mass sources.} 


   \maketitle
%
%%
%________________________________________________________________
\section{Introduction}
\label{sec.int}
The small molecules methylamine (CH$_{3}$NH$_{2}$) and hydroxylamine (NH$_{2}$OH) with an amine (-NH$_{2}$) functional group have both been suggested as precursors to the formation of amino acids \citep{Blagojevic2003,Holtom2005,Snow2007,Bossa2009,Barrientos2012,Garrod2013}. Reactions involving these molecules could explain the presence of the simplest amino acid glycine in comets \citep{Elsila2009,Altwegg2016}.
%These reactions can explain the formation of the simplest amino acid glycine and its presence in comets \citep{Elsila2009,Altwegg2016}. 
Despite their importance, both CH$_{3}$NH$_{2}$ and NH$_{2}$OH have turned out to be quite elusive molecules in the interstellar medium. CH$_{3}$NH$_{2}$ has exclusively been detected towards Sgr B2 and tentatively towards Orion KL \citep[e.g.][]{Kaifu1974,Pagani2017}. Upper limit abundances of CH$_{3}$NH$_{2}$ towards other high-mass sources are generally found to be consistent with values determined towards Sgr B2 \citep{Ligterink2015}. In the Solar System CH$_{3}$NH$_{2}$ has been detected in comets 81P/Wild 2 and 67P/Churyumov-Gerasimenko \citep[hereafter 67P/C-G][]{Elsila2009,Goesmann2015,Altwegg2017}. NH$_{2}$OH has not been detected thus far, down to upper limit abundances of $\sim$10$^{-11}$ with respect to H$_2$ \citep{Pulliam2012,Mcguire2015}.

The lack of detection of these two molecules not only constrains amino acid formation, but also contrasts with model predictions. \citet{Garrod2008} predicted efficient CH$_{3}$NH$_2$ formation from the radical addition reaction CH$_{3}$ + NH$_{2}$ in their models, whereas NH$_{2}$OH is assumed to form from the NH + OH addition followed by hydrogenation and NH$_{2}$ + OH reactions on ice surfaces. Abundances of CH$_{3}$NH$_{2}$ and NH$_{2}$OH are predicted to be on the order of 10$^{-6}$--10$^{-7}$, depending on the model. It is generally found that these models overproduce both molecules compared with observations \citep{Pulliam2012,Ligterink2015}. Therefore, other formation, reaction or destruction mechanisms need to be considered. 

Several laboratory experiments have investigated the formation of NH$_{2}$OH and CH$_{3}$NH$_2$. \citet{Zheng2010} show the formation of NH$_{2}$OH from electron irradiated H$_{2}$O:NH$_{3}$ ice mixtures, while \citet{He2015} produce the molecule by oxidation of NH$_{3}$ ice. Alternatively, NH$_{2}$OH is seen to efficiently form from the solid-state hydrogenation of nitric oxide \citep[NO;][]{Congiu2012a,Fedoseev2012}. In this scenario, NO is accreted from the gas-phase onto dust grains during cloud collapse \citep{Visser2011}. Nitrous oxide (N$_{2}$O) is found as a by-product of NO hydrogenation reactions.
%Both \citet{Fedoseev2012} and \citet{Congiu2012a} find nitrous oxide (N$_{2}$O) as a by-product of NO hydrogenation in a variety of experimental conditions.
%The precursor of NH$_{2}$OH, the molecule NO has been observed in a variety of sources \citep[e.g.,][]{Liszt1978,Mcgonagle1990,Ziurys1991,Watanabe2012,Yildiz2013}. 
NO has been observed in a variety of sources \citep[e.g.][]{Liszt1978,Yildiz2013,Codella2017}. It is thought to mainly form via the N + OH $\rightarrow$ NO + H neutral-neutral reaction in the gas-phase. Observations suggest that N$_2$O is related to NO \citep{Ziurys1994,Halfen2001}.

CH$_{3}$NH$_{2}$ formation has been demonstrated in electron irradiated CH$_{4}$:NH$_{3}$ ice mixtures \citep{Kim2011,Forstel2017}, with the main formation pathways suggested to proceed through CH$_{3}$ + NH$_{2}$ radical reactions. \citet{Theule2011} investigated hydrogenation of solid hydrogen cyanide (HCN) and methanimine (CH$_2$NH), both of which lead to CH$_3$NH$_2$ formation. CH$_{2}$NH is hypothesized to have a larger reaction probability than HCN and reaction pathways to CH$_{3}$NH$_{2}$ may be completely different for reactions starting from either HCN or CH$_{2}$NH.
%We focus on the precursor CH$_{2}$NH, NO and the side product N$_{2}$O{\bf This doesn't flow}. 
In contrast with CH$_{3}$NH$_{2}$, its potential precursor CH$_{2}$NH has been observed in numerous sources \citep{Dickens1997,Nummelin2000,Belloche2013,Suzuki2016}. \citet{Halfen2013} investigated the relationship between this molecule and CH$_{3}$NH$_{2}$ in Sgr B2 and concluded that the two species have different formation routes, due to observed differences in rotational temperature and distribution. Interestingly, CH$_{2}$NH has also been implied as a precursor to amino acid formation \citep[e.g.][]{Woon2002,Danger2011}.

Searches for CH$_{3}$NH$_{2}$ and NH$_{2}$OH have so far mainly focused on high-mass sources. Detections or upper limits of these two molecules and their potential precursors towards a low-mass source would therefore expand our understanding of amine-containing molecules and their formation in the ISM. The low-mass solar-type protostellar binary IRAS 16293--2422 (hereafter IRAS 16293) is an ideal source for such a study. Its physics and chemistry are well studied and it is abundant in complex organic molecules \citep[e.g.][]{jorgensen2016}. Abundance ratios will therefore constrain the chemistry of -NH$_{2}$ molecules as has been done for other nitrogen-bearing species \citep{Coutens2016,Ligterink2017}. In this paper we present the first detection of CH$_{2}$NH and N$_{2}$O towards a low-mass protostar. NO is also detected and analysed. The abundances of NH$_{2}$OH and CH$_{3}$NH$_{2}$ are constrained by upper limits from non-detections. 

%NO rapidly hydrogenated to NH$_2$OH at low ice temperatures Congiu et al. 2012 . . A critical parameter here is the competition of the different channels for reaction of HNO + H, which can either go back to NO + H$_2$ or form H$_2$NO

%These results for IRAS 4A suggest that a long pre-collapse                                   stage is characteristic of the earliest stages of star formation, in which atomic and molecular oxygen are frozen-out onto the dust grains and converted into water ice, as proposed by Bergin et al. (2000). Similarly, the rapid conversion of NO to other species on the grains limits its gas-phase abundance. It is clear that the grain surface processes are much more important than those of the pure gas-phase chemistry in explaining the O$_2$ and NO observations. The timescale for NGC 1333 IRAS 4A is at the long end of that inferred more generally from observations.
%__________________________________________________________________
\section{Observations and data analysis}
\label{sec.obs}
The observations were taken as part of the Protostellar Interferometric Line Survey (PILS), an unbiased spectral survey using the Atacama Large Millimeter Array (ALMA, \citealt{jorgensen2016}). The survey covers a spectral range of 329.147 to 362.896\,GHz in Band 7, obtained with the 12\,m array and the Atacama Compact Array (ACA). The combined data set analysed in this work was produced with a circular restoring beam of 0\farcs5. The maximum recoverable scale is 13$^{\arcsec}$. A spectral resolution of 0.2\,km s$^{-1}$ and a root mean square (RMS) noise level of about 7--10\,mJy beam$^{-1}$\,channel$^{-1}$, i.e. approximately 4--5\,mJy beam$^{-1}$\,km s$^{-1}$ is obtained. The dataset has a calibration uncertainty of 5--10\%.

The spectral analysis presented below is performed towards source B in IRAS 16293 at a position offset by one beam diameter (0\farcs5) from the continuum peak position in the south west direction ($\alpha_{J2000}$=16$^{\rm h}$32$^{\rm m}$22\fs58, $\delta_{J2000}$=$-$24$^{\circ}$28$^{\arcmin}$32.8$^{\arcsec}$). This position is used for most other PILS molecular identifications and abundance analyses \citep{Coutens2016,Lykke2017,Persson2017}. Lines are particularly narrow towards this position, only 1\,km s$^{-1}$, limiting line confusion.

The spectra are analysed with the CASSIS software\footnote{\url{http://cassis.irap.omp.eu/}} and linelists from the Jet Propulsion Laboratory (JPL\footnote{\url{http://spec.jpl.nasa.gov}}) catalog for molecular spectroscopy \citep{Pickett1998} and Cologne Database for Molecular Spectroscopy \citep[CDMS;][]{muller2001,muller2005}.  Specifically rotational spectroscopic data of NO, N$_{2}$O, CH$_{2}$NH, NH$_{2}$OH and CH$_{3}$NH$_{2}$ are used \citep{Kirchhoff1973,Pickett1979,Morino2000,Ilyusin2005,Ting2014}. Identified transitions are fitted with synthetic spectra, assuming Local Thermodynamic Equilibrium (LTE). The input for these fits is given by the column density ($N$), excitation temperature ($T_\mathrm{ex}$), line width (FWHM = 1~km~s$^{-1}$), peak velocity ($V_\mathrm{peak}$) and a source size of 0\farcs5, based on the observed spatial extent of the emission. The CASSIS software takes beam dilution into account by $\theta_{\rm source}^{\rm 2}$ / ($\theta_{\rm source}^{\rm 2} + \theta_{\rm beam}^{\rm 2}$).

Warm, dense dust is present around IRAS~16293B and its emission is coupled to the molecular line emission. This affects the line strength of molecular emission and needs to be corrected for by taking a higher background temperature into account, see Eq. \ref{eq:dust} for the line brightness $T_{\rm B}(\nu)$.

\begin{equation}
\label{eq:dust}
T_{\rm B}(\nu) = T_{\rm 0} \left(\frac{1}{e^{T_{\rm 0}/T_{\rm ex}}-1} - \frac{1}{e^{T_{\rm 0}/T_{\rm bg}}-1} \right)\left( 1-e^{-\tau\nu} \right),
\end{equation}

where $T_{\rm 0}$ = $h\nu$/$k_{\rm B}$, with $h$ being the Planck constant and $k_{\rm B}$ the Boltzmann constant. For the one beam offset position at source B $T_{\rm bg}$ = 21~K, compared to the usual cosmic microwave background temperature of $T_{\rm bg}$ = 2.7~K.

To compute the best-fit spectral models for each molecule, a grid of models is run to determine the best fitting synthetic spectra. For CH$_{2}$NH, $N$ was varied between 1$\times$10$^{14}$ -- 5$\times$10$^{15}$ cm$^{-2}$ and $T_\mathrm{ex}$ between 25 -- 300~K, while for NO and N$_{2}$O the parameter space of $N$ = 1$\times$10$^{15}$ -- 1$\times$10$^{17}$ cm$^{-2}$ and $T_\mathrm{ex}$ = 25 -- 400~K was explored. The $V_\mathrm{peak}$ parameter space is explored between 2.5 -- 2.9\,km s$^{-1}$ and the line FWHM is fixed to 1\,km s$^{-1}$, as found for most transitions at this position in the PILS data set \citep{jorgensen2016}. From this model grid, we obtain the best-fit model solutions through $\chi^2$ minimisation. Generally, we accept fits that are within the observational uncertainty of the observed line profile as judged by eye.

The uncertainty in the data derives from the calibration error (10\%) and the RMS noise (mJy~beam$^{-1}$\,channel$^{-1}$) as $\sqrt{(0.1T_{\rm peak})^{2}+(RMS)^{2}}$, with $T_{\rm peak}$ the peak brightness of a line. This uncertainty also applies to any best-fit synthetic model. In Table \ref{tab.lines} in Appendix \ref{ap:spec}, model parameters for each detected transition as well as their errors are given. The error on the peak velocity equals the channel width of the observations. The FWHM is kept as a fixed parameter. The error on the peak brightness temperature ($T_{\rm peak}$) is determined from the combination of the RMS and the calibration error of the data, as mentioned above. The largest source of error on the column density and excitation temperature comes from the quality of the fit of the spectral model to the data, particularly due to the large number of blended lines in the PILS dataset. The best-fit model results are listed in Table \ref{tab:coldens}.

For the non-detected species the formalism in \citet{Ligterink2015} is used to derive upper limit column densities based on the 3$\sigma$ upper limit line intensities of the strongest transitions in line-free ranges of the PILS data. The 1$\sigma$ limit is given by 1.1$\cdot \sqrt{\delta\nu \cdot FWHM} \cdot$RMS, where the factor 1.1 accounts for a 10\% calibration uncertainty, $\delta\nu$ the velocity resolution of the data (0.2~km~s$^{-1}$), FWHM is the line width at the one beam offset position in source B (1\,km\,s$^{-1}$) and RMS the noise in mJy~beam$^{-1}$\,bin$^{-1}$.

%__________________________________________________________________
\section{Results}
\label{sec.res}

\begin{table*}
\footnotesize
\caption{Column densities and rotational temperatures at the one beam offset position around source B from the ALMA-PILS data \label{tab:coldens}}
\center
\begin{tabular}{llccc}
\hline
\hline
Molecule&Chemical formula&$N_\mathrm{tot}^{\dagger}$ (cm$^{-2}$)&$T_\mathrm{ex}$ (K) & $V_{\rm peak}$ (km\,s$^{-1}$)\\
\hline
Nitric oxide&NO&\phantom{$\le$}(1.5 -- 2.5)$\times$10$^{16}$&40 -- 150& 2.5 $\pm$0.2\\
Nitrous oxide&N$_2$O&$\geq$4.0$\times$10$^{16,\ddagger}$&25 -- 350& 2.5 $\pm$0.2\\ 
Hydroxylamine&NH$_2$OH&$\leq$3.7$\times$10$^{14}$&100 & 2.7 \\
Methanimine&CH$_2$NH&\phantom{$\le$}(6.0 -- 10.0)$\times$10$^{14}$&70 -- 120& 2.7 $\pm$0.2\\
Methylamine&CH$_3$NH$_2$&$\leq$5.3$\times$10$^{14}$&100  & 2.7 \\
\hline
\end{tabular}
\\
\tablefoot{All models assume LTE, a FWHM of 1\,km\,s$^{-1}$ and a source size of 0\farcs5. $^{\dagger}$Upper limits are 3$\sigma$ and determined for $T_{\rm ex}$ = 100~K. $^{\ddagger}$Due to the large uncertainty on the highest excitation temperature only the lower limit column density is given.}
%\\$^{\dagger}$ Difference in velocity peak from $V_\mathrm{LSR}$ due to error in rest frequencies. 
\end{table*}

\begin{figure}
 \begin{center} 
 \includegraphics[width=0.5\hsize]{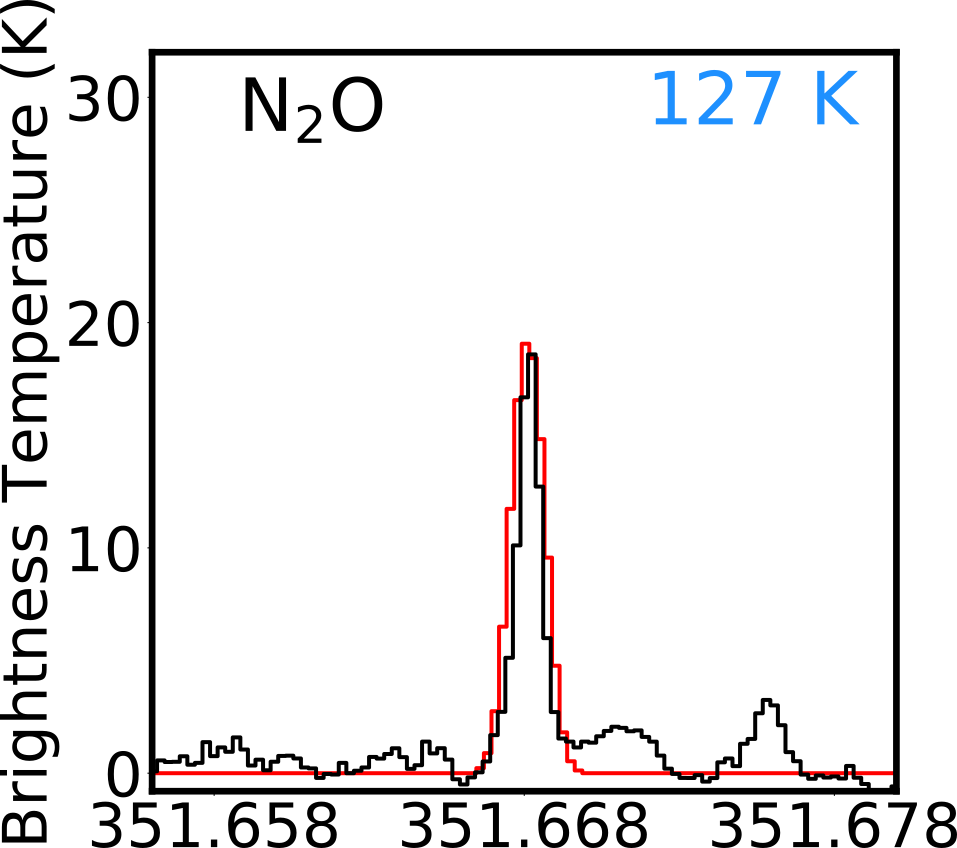}
 \end{center} 
 \caption{The N$_{2}$O transition identified in the PILS spectrum at one beam offset around source B (black), with the synthetic spectrum for $T_{\rm ex}$ = 100~K and $N$ = 5.0$\times$10$^{16}$ cm$^{-2}$ (red). The upper state energy of the transition is indicated in blue.
 \label{fig:lines3}} 
 \end{figure}

Spectral lines from CH$_2$NH, NO and N$_2$O emission are detected, while those of CH$_{3}$NH$_{2}$ and NH$_{2}$OH are not. An overview of the detected transitions is presented in Table \ref{tab.lines} in Appendix A. Figures \ref{fig:lines1}, \ref{fig:lines2} and \ref{fig:lines3} show the detected transitions of CH$_{2}$NH, NO and N$_{2}$O towards source B, with synthetic spectra overplotted.

 %%%%%%%%%%%%%%%%%%%%%%%%%%%%%%%%%%%%%%%%%%%%%%
 \begin{figure*}
 \begin{center} 
 \includegraphics[width=\hsize]{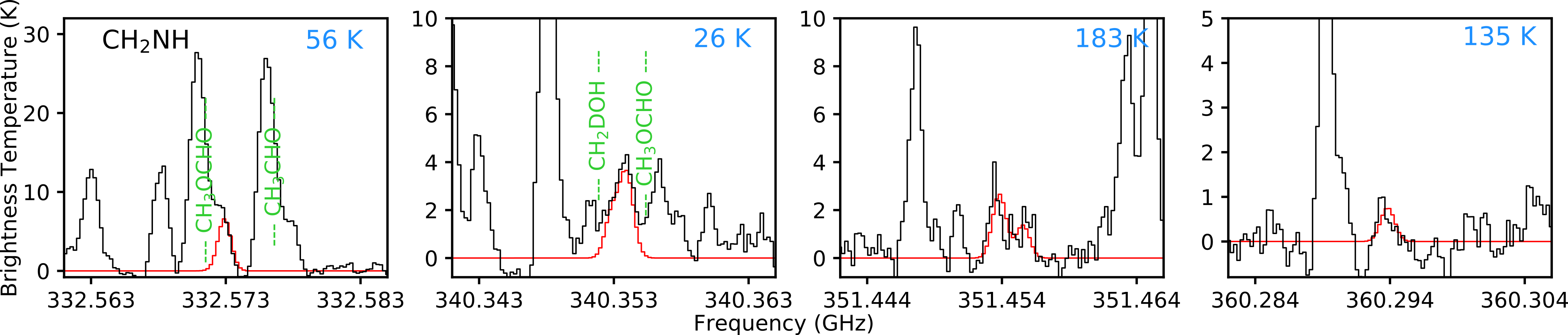}
 \end{center} 
 \caption{All detected transitions of CH$_{2}$NH in the PILS spectrum at one beam offset around source B (black), with the synthetic spectrum for $T_{\rm ex}$ = 100~K and $N$ = 8.0$\times$10$^{14}$ cm$^{-2}$ (red) and other known species indicated in green. The upper state energy of each CH$_{2}$NH transition is indicated in blue.
 \label{fig:lines1}} 
 \end{figure*} 
 %%%%%%%%%%%%%%%%%%%%%%%%%%%%%%%%%%%%%%%%%%%%%%

CH$_{2}$NH is detected for the first time towards a low-mass protostellar source. A total of 18 hyperfine transitions in four different spectral features are detected. The first feature shows up at 332.575 GHz and is blended with a neighbouring methylformate (CH$_{3}$OCHO) line. At 340.354 GHz five hyperfine lines form a distinct feature that is unblended with any other molecule. Around 351.455 GHz a distinct double peak feature is found and at 360.294 GHz the fourth feature is seen. These last three features are found to be unblended. ALMA observations in bands 3, 4, 6 and 9 towards IRAS 16293 were used to search for additional lines, but spectral features of CH$_{2}$NH fall outside the spectral windows of these observations. Publicly available data of the single dish TIMASSS survey towards IRAS 16293 \citep{Caux2011} were analysed for CH$_{2}$NH spectral features, but no lines were detected, suggesting that the emission indeed arises mostly from a compact source.

Upper state energies ($E_\mathrm{up}$) of the detected features range from 26 to 183~K, making it possible to constrain the excitation temperature. Using a grid of models, the emission of this species can be fitted with excitation temperatures ranging between 70--120~K and column densities of 6.0$\times$10$^{14}$--10.0$\times$10$^{14}$ cm$^{-2}$. Outside this temperature range synthetic spectra cannot reproduce the observed spectrum, as can be seen in Fig.~\ref{fig:Tex_CH2NH} in Appendix~\ref{ap:add_spec} for fits at $T_{\rm ex}$ = 50 and 150~K. The peak velocity is found to be 2.7~km~s$^{-1}$. For the best fits, no anti-coincidence is found between the synthetic and observed spectrum. Figure \ref{fig:lines1} shows the synthetic spectrum at $T_{\rm ex}$ = 100~K and $N$ = 8.0$\times$10$^{14}$ cm$^{-2}$.

%A fixed $T_{\rm ex}$ of 100~K is used for all molecules to determine their best fit column densities. The emission of the CH$_{2}$NH transitions is well fitted with a column density of 8.0(2.0)$\times$10$^{14}$ cm$^{-2}$, where uncertainties due to differences in excitation temperature are taking into account.

 %%%%%%%%%%%%%%%%%%%%%%%%%%%%%%%%%%%%%%%%%%%%%% 
 \begin{figure*}
 \begin{center} 
 \includegraphics[width=\hsize]{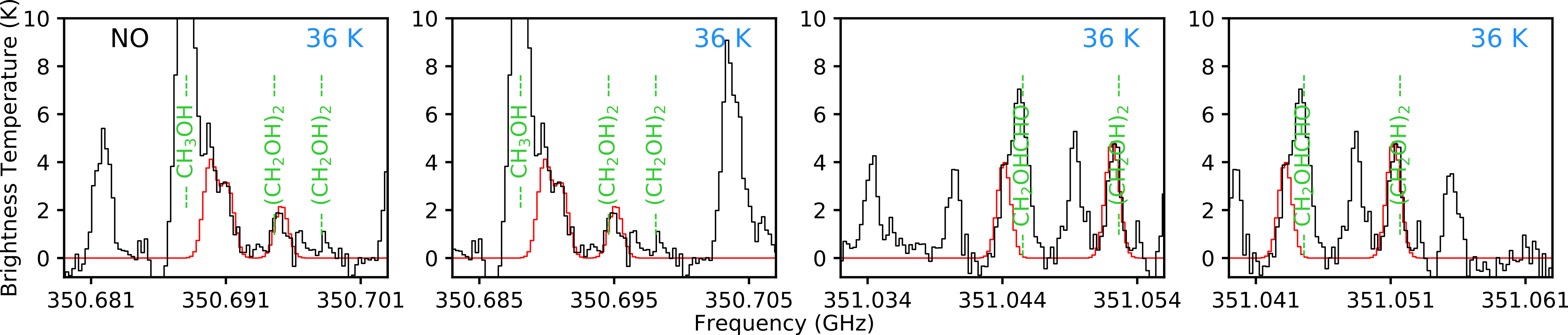}
 \end{center} 
 \caption{All detected transitions of NO in the PILS spectrum at one beam offset around source B (black), with the synthetic spectrum for $T_{\rm ex}$ = 100~K and $N$ = 2.0$\times$10$^{16}$ cm$^{-2}$ (red) and other known species indicated in green. The upper state energy of each NO transition is indicated in blue.
 \label{fig:lines2}} 
 \end{figure*}  
 %%%%%%%%%%%%%%%%%%%%%%%%%%%%%%%%%%%%%%%%%%%%%%

Five NO transitions, each with an upper state energy of 36~K, are detected. A number of NO lines at $E_{\rm up}$ = 209~K are in the spectral range covered by the data, but are not detected. The excitation temperature can therefore be constrained to be lower than 150~K, otherwise anti-coincidences with the $E_{\rm up}$ = 209~K lines in the synthetic spectrum show up, as can be seen in Fig.~\ref{fig:Tex_NO} in Appendix~\ref{ap:add_spec}. The emission can be fitted with $T_{\rm ex}$ = 40 -- 150~K and $N$ = 1.5$\times$10$^{16}$ -- 2.5$\times$10$^{16}$ cm$^{-2}$. A peak velocity of $V_\mathrm{peak}$ = 2.5\,km\,s$^{-1}$ is found for the NO lines, slightly offset from the $V_\mathrm{LSR}$ = 2.7\,km\,s$^{-1}$ of source B. A similar offset is seen for other species in the PILS dataset, specifically acetaldehyde and ethylene oxide \citep{Lykke2017,Jorgensen2017}. Note that seven NO lines have already been detected towards the same source in the TIMASSS survey by \citet{Caux2011}. These lines do not arise from the same regions. The lines from the TIMASSS survey are emitted in the cold envelope and source A, since they show a $V_\mathrm{\rm LSR}$ of $\sim$ 4.0\,km\,s$^{-1}$ \citep{Caux2011}, while the lines in the PILS data come from the hot core region in source B. Figure \ref{fig:lines2} shows the synthetic model to the observational data at an excitation temperature of 100\,K and column density of 2.0$\times$10$^{16}$ cm$^{-2}$.

Three N$_{2}$O transitions fall in the frequency range of the PILS data, one $\varv$=0 line with $E_{\rm up}$ = 127~K and two $\varv_{2}$=1 lines with $E_{\rm up}$ = 973~K. In the observed spectrum a feature is found that fits the $\varv$=0 line at 351.667 GHz. Since no other known species were found to correspond to this feature, a tentative detection of N$_{2}$O can be claimed. Similar to NO, $V_\mathrm{peak}$ = 2.5 km s$^{-1}$ fits the position of this line. Based on the non-detection of the $\varv_{2}$=1 lines, the excitation temperature can be constrained to be lower than 350~K.

%At $T_{\rm ex}$ = 100~K a column density of 5.0$\times$10$^{16}$ cm$^{-2}$ for N$_{2}$O is derived. 
%, with an uncertainty of $\pm$1.0$\times$10$^{16}$ cm$^{-2}$ over the entire possible temperature range.

\begin{figure}
\begin{center} 
\includegraphics[width=9cm, angle=0, clip =true, trim = 1.5cm 5.2cm 1.5cm 5.2cm]{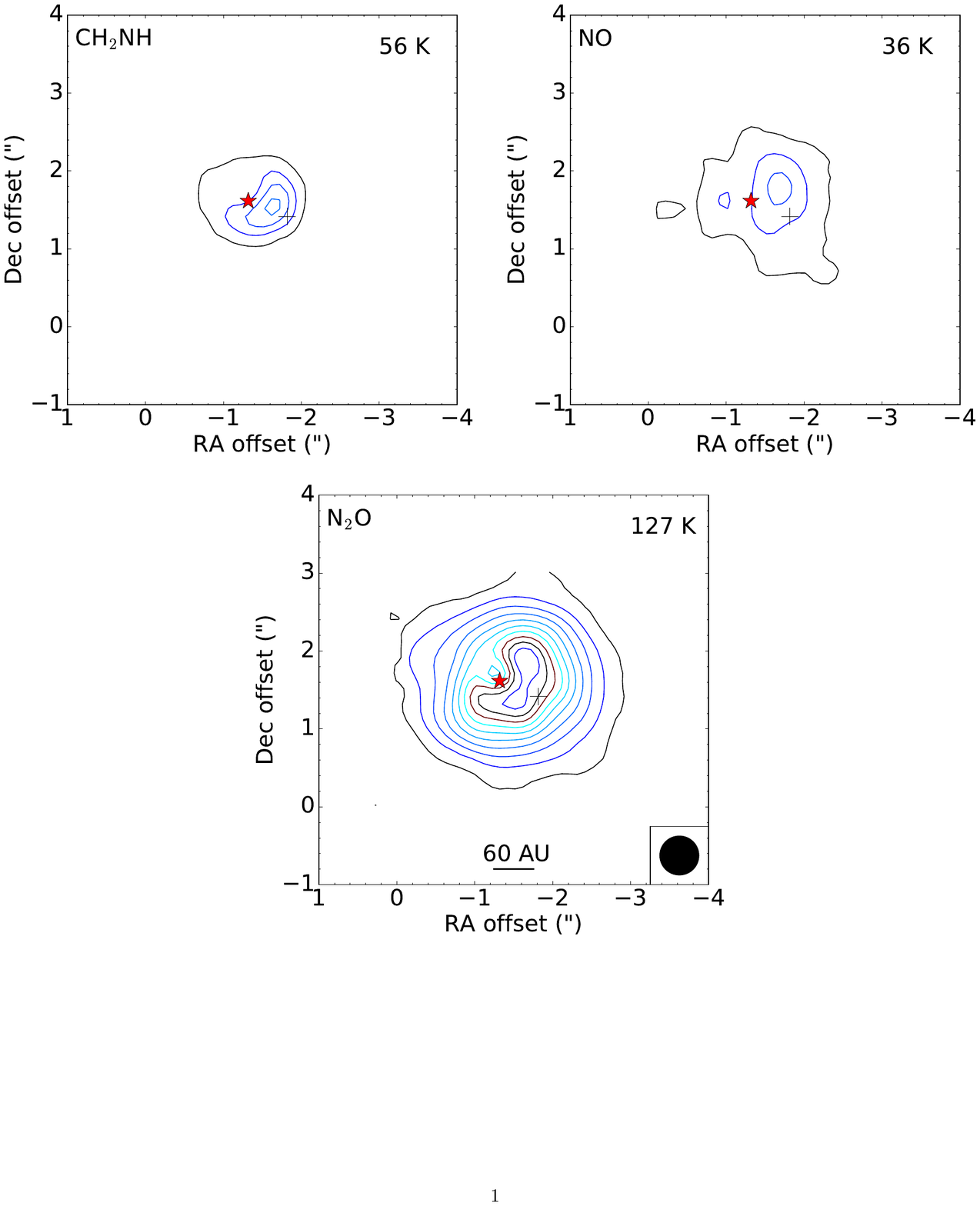}
\end{center} 
\caption{Integrated emission maps of the 332.572\,GHz line of CH$_2$NH, the 351.052\,GHz line of NO, and the 351.668\,GHz line of N$_2$O. The emission is integrated between 2.2 and 3.2\,km\,s$^{-1}$. The axes show the position offset from phase centre of the observations. Contour levels start at 30 mJy\,kms$^{-1}$ and increase in steps of 45 mJy\,kms$^{-1}$. The red star marks the peak continuum position and the black cross marks the one beam offset position where the spectra are analysed. \label{fig:mapcomp}} 
\end{figure}

To further support this assignment, additional datasets were checked. No N$_{2}$O lines were covered in ALMA Band 3, 4, 6 and 9 datasets towards IRAS 16293. However, analysis of data from the TIMASSS survey resulted in the identification of a number of features. Four N$_{2}$O lines are found to be unblended (see Fig. \ref{fig:timasss_n2o} in Appendix \ref{ap:timasss_n2o}). Based on Gaussian fitting of the profiles, the peak velocity of the four lines is 2.5\,km\,s$^{-1}$, similar to the N$_{2}$O transition detected towards source B in the PILS survey. However, velocity components at 3.9\,km\,s$^{-1}$, resulting from source A or extended emission, cannot be entirely ruled out. Based on a comparison of the N$_{2}$O 351.667 GHz line flux of 394\,mJy beam$^{-1}$\,km s$^{-1}$ or 15.7\,K\,km s$^{-1}$ in the ALMA-PILS and 0.2\,K\,km s$^{-1}$ in the JCMT-TIMASSS data, the emitting area in the TIMASSS data is found to be around 1\farcs6. This is supported by the integrated emission map of the N$_{2}$O 351.667 GHz line (Fig. \ref{fig:mapcomp}), where the emission has a larger FWHM than other species. The best fit to the optically thin lines in the TIMASSS data is found for $T_{\rm ex}$ = 68~K and $N$ = 1.7$\times$10$^{16}$ cm$^{-2}$ (See Fig. \ref{fig:rtd_n2o}). Note that the precise source size for N$_{2}$O can change the excitation temperature by a couple tens of Kelvin, but can be constrained to be below 100~K, and column density by $\pm$25\%. See Appendix \ref{ap:timasss_n2o} for further details.

In the PILS data, at $T_{\rm ex}$ = 100~K a column density of 5.0$\times$10$^{16}$ cm$^{-2}$ is derived for the N$_{2}$O 351.667 GHz transition, see Fig. \ref{fig:lines3}. However, since the TIMASSS data indicate that the excitation temperature is likely lower than 100~K, this line was fitted with $T_{\rm ex}$ = 25--100~K. For these temperatures, emission is found to be optically thick and the column density is constrained to be $\geq$5.0$\times$10$^{16}$ cm$^{-2}$ for the entire temperature range. Excitation temperatures between 100 and 350~K cannot be neglected and for this range column densities of $\geq$4.0$\times$10$^{16}$ cm$^{-2}$ are derived.

Finally, lines of the $^{15}$N and $^{18}$O isotopologues of N$_{2}$O were searched for in the PILS data, but not detected to levels above the standard $^{14}$N/$^{15}$N and $^{16}$O/$^{18}$O ISM ratios \citep[see][]{Wilson1999,Milam2005}. Both the $^{15}$N$^{14}$NO and $^{14}$N$^{15}$NO transition are blended with an unknown feature and a HN$^{13}$CO line, respectively. For N$_{2}^{18}$O an upper limit column density of $\leq$1$\times$10$^{15}$ cm$^{-2}$ at $T_\mathrm{ex}$ = 100~K is determined, resulting in N$_{2}$O/N$_{2}^{18}$O > 40 compared to an expected $^{18}$O/$^{16}$O ratio of 560 \citep{Wilson1999}.

%Based on the non-detection of the $\varv_{2}$=1 lines, the excitation temperature can be determined to be lower than 250~K. At $T_{\rm ex}$ = 100~K a column density of 5.0$\times$10$^{16}$ cm$^{-2}$ for N$_{2}$O is derived, with an uncertainty of $\pm$1.0$\times$10$^{16}$ cm$^{-2}$ over the entire possible temperature range.}

 %%%%%%%%%%%%%%%%%%%%%%%%%%%%%%%%%%%%%%%%%%%%%%
 %%%%%%%%%%%%%%%%%%%%%%%%%%%%%%%%%%%%%%%%%%%%%%

Figure \ref{fig:mapcomp} shows the emission maps of the 332.572\,GHz line of CH$_2$NH, the 351.052\,GHz line of NO, and the 351.668\,GHz line of N$_2$O towards source B. Emission is generally found to be compact and compares well with emission maps of other molecules towards source B, such as NH$_{2}$CHO \citep{Coutens2016}. Emission of N$_2$O is more extended than that of NO.

Transitions of both CH$_{3}$NH$_{2}$ and NH$_{2}$OH are not detected and therefore upper limit column densities are determined. For CH$_{3}$NH$_{2}$, the 6$_{1}$ $\rightarrow$ 5$_{0}$ transition at 357.440\,GHz best constrains the upper limit column density. For a 3$\sigma$ upper limit line intensity of 23\,mJy\,km\,s$^{-1}$, the upper limit column densities versus the rotational temperatures are plotted in Fig.~\ref{fig:ch3nh2_up} in Appendix \ref{ap:non}. At $T_{\rm ex}$ = 100~K, the upper limit column density is $\leq$5.3$\times$10$^{14}$ cm$^{-2}$. 

NH$_2$OH has three strong transitions at 352.522, 352.730 and 352.485\,GHz for the 7$_{0}$ $\rightarrow$ 6$_{0}$, 7$_{1}$ $\rightarrow$ 6$_{1}$, 7$_{2}$ $\rightarrow$ 6$_{2}$ transitions, respectively. Slightly different RMS noise conditions apply around each of these transitions, resulting in a 3$\sigma$ of 27, 21 and 27\,mJy\,km\,s$^{-1}$, respectively. The upper limit column density versus rotational temperature plot is shown in Fig.~\ref{fig:nh2oh_up} in Appendix \ref{ap:non}. The transition at 352.522 \,GHz constrains the column density most, resulting in an upper limit column density of NH$_{2}$OH of $\leq$3.7$\times$10$^{14}$ cm$^{-2}$ at $T_{\rm ex}$ = 100~K. 

Table \ref{tab:coldens} lists the column densities and excitation temperatures for the five molecules under investigation in this work. For the three detected species the column density and excitation temperature ranges that can fit the emission are listed. For the non-detected species CH$_{3}$NH$_{2}$ and NH$_{2}$OH the upper limit column density is determined for $T_{\rm ex}$ = 100~K. The typical uncertainty of the column densities of these upper limits for a reasonable range of excitation temperatures is a factor of $\sim$2 (see Appendix \ref{ap:non}).
  
\begin{table*}
     \caption[]{Comparison of NO and N$_{2}$O abundances at one beam offset around source B.}
         \label{tab.no_n2o}
         $$
         \begin{tabular}{l c c c l}
            \hline
            \hline
            
            Source & NO/H$_2$ & N$_2$O/H$_2$ &  NO/N$_2$O & Reference \\
            
            \hline
            
          	IRAS16293--2422B	& $\leq$1.7$\times$10$^{-9,a}$	& $\sim$3.3$\times$10$^{-9,b}$	& $\leq$0.5 & This work	\\
            
           	\hline
           
            Sgr B2(M) & $\sim$9$\times$10$^{-9}$ & $\sim$1$\times$10$^{-9}$ & $\sim$10 & \citet{Ziurys1994} \\
   			Sgr B2(N) & 1.32$\times$10$^{-8}$ & 1.5$\times$10$^{-9}$ & 8.8 & \citet{Halfen2001} \\
            
            \hline
            
            Sgr B2 & 1$\times$10$^{-8}$ & -- & -- & \citet{Liszt1978} \\
            L134N & 6$\times$10$^{-8}$ & -- & -- & \citet{Mcgonagle1990} \\
            NGC 1333 IRAS 4A, & 2.3$\times$10$^{-8}$ & -- & -- & \citet{Yildiz2013} \\
            L1157-B1 & 4--7$\times$10$^{-7}$ & -- & -- & \citet{Codella2017} \\
            SVS13-A & $\leq$3$\times$10$^{-7}$ & -- & -- & \citet{Codella2017} \\

            \hline
            
         \end{tabular}     
         $$     
		\tablefoot{$^{a}$Ratio determined from a lower limit H$_{2}$ column density. $^{b}$Ratio determined from lower limit N$_{2}$O and H$_{2}$ column densities.}
\end{table*}

\begin{table*}[ht!]
\center
\footnotesize
     \caption{Comparison of CH$_{3}$NH$_{2}$ abundances at one beam offset around source B. \label{tab.ch3nh2}}
         \begin{tabular}{lcccl}
            \hline
            \hline
            
            Source & CH$_{3}$NH$_{2}$/CH$_{2}$NH & CH$_{3}$NH$_{2}$/NH$_{2}$CHO &  CH$_{3}$NH$_{2}$/CH$_{3}$OH & Reference \\
            
            \hline
            
          	IRAS 16293--2422B	& $\leq$0.88	& $\leq$0.053	& $\leq$5.3$\times$10$^{-5}$ & This work	\\
			
           	\hline
            
			Sgr B2 & $\sim$1 & 0.57 & 0.017 & \citet{Turner1991} \\
			Sgr B2(M) & 31 & 3.2 & 0.008 & \citet{Belloche2013}  \\
           	Sgr B2(N) & 0.75 & 0.43 & 0.034 & \citet{Belloche2013} \\
           	Sgr B2(N) & 5.5	& -- & -- & \citet{Halfen2013}	\\
           	Sgr B2(N) & 7.1 & 2.1 & 0.1 & \citet{Neill2014}  \\  
            
            \hline
            
            Hot core model & -- & 1.1--1.7 & 0.034--0.13 & \citet{Garrod2008}$^{a}$ \\
            Hot core model & 0.5--7.3 & 0.2--2.3 & 1.8--7.3$\times$10$^{-3}$ & \citet{Garrod2013}$^{a}$\\
                     
           	\hline  
            
         \end{tabular}     
              
	\tablefoot{$^{a}$Abundance ranges taken from the F(ast), M(edium) and S(low) warm-up models.}
\end{table*}

\begin{table*}[!t]
\centering
         \footnotesize
 \caption{Comparison of CH$_2$NH abundances at one beam offset around source B. \label{tab.ch2nh}}
         \begin{tabular}{lcccl}

            \hline
            \hline
            
            Source & CH$_{2}$NH/NH$_{2}$CHO &  CH$_{2}$NH/CH$_{3}$OH & CH$_{2}$NH/H$_{2}$ & Reference \\
            
            \hline
            
          	IRAS 16293--2422B	& 0.08	& 8.0$\times$10$^{-5}$ & $<$6.7$\times$10$^{-11}$ & This work	\\

           	\hline
            
			Sgr B2 		& $\sim$0.57 & $\sim$0.017 & -- & \citet{Turner1991}\\
			Sgr B2(M) & 0.10 & 5.5$\times$10$^{-4}$ & -- & \citet{Belloche2013}  \\
           	Sgr B2(N) & 0.57 & 0.044 & -- & \citet{Belloche2013} \\
			Sgr B2(N) & -- & -- & 3.0$\times$10$^{-10}$ & \citet{Halfen2013}\\          
           	Sgr B2(N) & 0.30 & 0.014 & 8.8$\times$10$^{-9}$ & \citet{Neill2014} \\
         	Orion KL & -- & 1.9$\times$10$^{-3}$   & 4.2$\times$10$^{-9}$ & \citet{Crockett2014} \\                 
            \hline
            Hot core model & 0.03--5.17 & 1.0--3.9$\times$10$^{-3}$ & 6.8--80$\times$10$^{-9}$ & \citet{Garrod2013}$^{a}$\\
            \hline
         \end{tabular}     
	\tablefoot{$^{a}$Abundance ranges taken from the F(ast), M(edium) and S(low) warm-up models.} 
\end{table*}

%__________________________________________________________________
\section{Astrochemical implications}
\label{sec.dis}

To put the results into context, the column density ratios between the molecules studied here and formamide (NH$_{2}$CHO), methanol (CH$_{3}$OH) and molecular hydrogen (H$_{2}$) are computed and compared with other sources. The column densities of these reference species towards the same, one beam offset position in source B are: NH$_{2}$CHO = 1$\times$10$^{16}$, CH$_{3}$OH = 1$\times$10$^{19}$ and H$_{2}$ $>$1.2 $\times$10$^{25}$ cm$^{-2}$ \citep{Coutens2016,jorgensen2016}. The H$_{2}$ lower limit column density is determined towards the continuum peak position of source B, but the same lower limit should hold for the one beam offset position being analysed in this paper, where the dust emission is still optically thick. Earlier results for Orion KL and for some Sgr B2 studies, used to put our ratios into context, are based on single dish data, whereas in this work interferometric data are used. It is important to stress that differences in abundance ratios do not necessarily reflect chemical differences, but could arise from the fact that single dish observations generally probe larger spatial scales and are more affected by beam dilution \citep[see][]{jorgensen2016}.

The NH$_{2}$OH upper limit abundance of $N$(NH$_{2}$OH)/$N$(H$_{2}$) $\leq$3.1$\times$10$^{-11}$ found in this work is comparable to upper limit abundances found for high-mass sources by \citet{Pulliam2012}, but significantly lower than the ice abundances of 7$\times$10$^{-9}$ $N$(H+H$_{2}$) found with a dark cloud model in \citet{Fedoseev2012}. One explanation could be that the reaction barriers in this model are too low or that destruction or competing pathways are missing. It could also be that the dominant pathway to NH$_{2}$OH is a gas-phase formation route, although such a route would likely also be hindered by a large reaction barrier or subject to efficient destruction or competing pathways. The more extensive gas-grain model of \citet{Garrod2013} also overpredicts this species, although the predicted abundance is lower by at least an order of magnitude. The full comparison of the hydroxylamine upper limit abundance compared with literature values is given in Table \ref{tab.nh2oh}.
One example of missing destructions reactions is given by laboratory experiments which show that thermal processing of NH$_{2}$OH:H$_{2}$O mixtures results in the conversion of NH$_{2}$OH into HNO, NH$_{3}$ and O$_{2}$ before the onset of desorption \citep{Jonusas2016}. Also, UV processing of ice mixtures containing NH$_{2}$OH results in the efficient destruction of this molecule \citep{Fedoseev2016}. The low gas-phase abundance of NH$_2$OH limits its involvement in gas-phase production routes of amino acids.

The NO abundance in IRAS\,16293B ($N$(NO)/$N$(H$_{2}$) $\leq$1.7$\times$10$^{-9}$, given as an upper limit due to the lower limit H$_{2}$ column density) is low compared to observational and modelling studies of dark clouds, where NO abundances of 10$^{-8}$--10$^{-6}$ are found \citep[e.g.][]{Mcgonagle1990,Visser2011}. The IRAS\,16293B abundance of NO is much more stringent than the NO upper limit of $\leq$3$\times$10$^{-7}$ derived towards the envelope of SVS13-A and also substantially lower than the NO abundance of (4--7)$\times$10$^{-6}$ towards the L1157-B1 shock \citep{Codella2017}.
%Observational and modelling studies have found NO to be efficiently produced in the gas-phase, resulting in dark cloud abundances of 10$^{-8}$--10$^{-6}$ \citep[e.g.,][]{Mcgonagle1990,Visser2011}. This contrasts with the low abundance of NO in IRAS\,16293 ($N$(NO)/$N$(H$_{2}$) $\leq$5.0$\times$10$^{-9}$). 
%This can indicate three things; 1) NO was not abundantly present in the dark cloud that IRAS 16293 formed from; 2) NO is depleted by hydrogenation and energetic processes in the ice (but the product NH$_{2}$OH does not release to the gas-phase); 3) gas-phase reactions destroy or convert NO. 
Modelling shows that NO is readily lost in the ice by conversion to other species, mainly NH$_{2}$OH \citep{Yildiz2013}, and in the gas-phase by photodissociation reactions in the hot core \citep{Visser2011} and thus could explain the depletion of NO in IRAS\,16293B.
The high N$_{2}$O abundance may indicate other loss pathways of NO as well. N$_{2}$O is found in laboratory ice experiments as a side product of NO hydrogenation and UV irradiation \citep{Congiu2012a,Fedoseev2012,Fedoseev2016} and suggested to form via the reaction NO + NH in warm gas \citep{Halfen2001}. The $N$(NO)/$N$(N$_{2}$O) $\leq$ 0.5 ratio hints to a scenario where NO is at least partly converted to N$_{2}$O in the ice and/or gas surrounding IRAS\,16293B. Interestingly the abundance ratio in IRAS 16293B contrasts with other known abundances toward Sgr B2, which have $N$(NO)/$N$(N$_{2}$O) $\sim$ 10 \citep{Ziurys1994,Halfen2001}, see Table \ref{tab.no_n2o}.

%In both cases, the high reactivity of NO can result in many efficient pathways to destroy this molecule. Chemical modelling of all three options in IRAS\,16293 is needed to investigate the details of how these abundances arise. 
%The high abundance of N$_{2}$O may hold clues for options 2 and 3 as well. In the laboratory ice experiments N$_2$O is found to be a side product of NO hydrogenation and VUV irradiation \citep{Congiu2012a,Fedoseev2012,Fedoseev2016} and could thus be an explanation for the presence of this molecule. However, in the case of hydrogenation N$_{2}$O is formed through NO dimers, abundantly present in the laboratory experiments. The presence of NO dimers in interstellar ices will be a lot lower and therefore unlikely to abundantly produce N$_{2}$O. \citet{Halfen2001} state that N$_2$O is formed from the reaction NO + NH in warm gas and could mean NO is readily lost in the IRAS 16293 hot core. In both cases the high reactivity of NO can result in many efficient pathways to destroy this molecule. Chemical modelling of all three options in IRAS 16293 is needed to investigate how these abundances exactly arise.

Table \ref{tab.ch3nh2} lists the abundance ratios of CH$_{3}$NH$_{2}$ in IRAS\,16293B, Sgr B2 and hot core models by \citet{Garrod2008} and \citet{Garrod2013}. The abundance ratios of CH$_{3}$NH$_{2}$/CH$_{2}$NH are slightly lower in IRAS\,16293B compared to Sgr B2 and the models of \citet{Garrod2013}, but ratios with respect to NH$_{2}$CHO and CH$_{3}$OH are lower by at least one to two orders of magnitude. Abundances of CH$_{2}$NH relative to NH$_{2}$CHO, CH$_{3}$OH and H$_{2}$ are given in Table \ref{tab.ch2nh}. The observed CH$_{2}$NH/NH$_{2}$CHO abundance ratio in IRAS\,16293B is lower than that in Sgr B2, but in range of model predictions. This can indicate that NH$_{2}$CHO is overpredicted in the models. The ratios with respect to CH$_{3}$OH and H$_{2}$ are in most cases at least an order of magnitude lower in IRAS\,16293B compared to Sgr B2 and models. The comparison of these abundances indicates that the formation of both CH$_{3}$NH$_{2}$ and CH$_{2}$NH is less efficient in the low-mass source IRAS\,16293B and overpredicted in models.

%For example, the UV flux is expected to be lower, which can result in a lower production rate of CH$_{3}$ and NH$_{2}$ radicals which are needed to form CH$_{3}$NH$_{2}$. 
CH$_{3}$NH$_{2}$ was mass spectrometrically detected on the comet 67P/C-G \citep{Goesmann2015}. However, the recent detection with the ROSINA-DFMS instrument indicates that it is present at a lower abundance, below the 1 percent level with respect to water, than initially thought \citep{Altwegg2017}, resulting in a CH$_{3}$NH$_{2}$/H$_{2}$O abundance of 1.2$\times$10$^{-4}$. Since IRAS\,16293B is assumed to resemble an early formation stage of our solar system, the low abundance of CH$_{3}$NH$_{2}$ in 67P/C-G and non-detection in IRAS\,16293B suggest that CH$_{3}$NH$_{2}$ is not efficiently formed in these environments. Furthermore, the low abundance of CH$_{3}$NH$_{2}$ limits the relevance of amino acid formation routes involving this species. Conversely, the detection of CH$_{2}$NH makes amino acid formation routes involving this molecule a more likely possibility.

%This letter presents the first detection of CH$_{2}$NH towards a low-mass protostar, IRAS 16293--2422. It is therefore a possible precursor to glycine as found in comets in the solar system. The abundance ratio between NO and N$_{2}$O indicates that NO is efficiently destroyed, either in ices or in the gas-phase. NH$_{2}$OH and CH$_{3}$NH$_{2}$ are not detected but upper limit abundances are determined. The non-detection of both these molecules constrains their possible involvement in the formation of amino acids. The low upper limit abundance of CH$_{3}$NH$_{2}$ reflects a less efficient formation mechanism in low mass sources. 

\section*{Acknowledgements}

We would like to thank the anonymous referee for the useful inputs given to this paper. This letter makes use of the following ALMA data: ADS/JAO.ALMA\#2013.1.00278.S. ALMA is a partnership of ESO (representing its member states), NSF (USA) and NINS (Japan), together with NRC (Canada) and NSC and ASIAA (Taiwan), in cooperation with the Republic of Chile. The Joint ALMA Observatory is operated by ESO, AUI/NRAO and NAOJ. Astrochemistry in Leiden is supported by the European Union A-ERC grant 291141 CHEMPLAN, by the Netherlands Research School for Astronomy (NOVA) and by a Royal Netherlands Academy of Arts and Sciences (KNAW) professor prize. The group of J.K.J. acknowledges support from a Lundbeck Foundation Group Leader Fellowship, as well as the ERC under the European Union's Horizon 2020 research and innovation programme through ERC Consolidator Grant S4F (grant agreement No 646908). Research at the Centre for Star and Planet Formation is funded by the Danish National Research Foundation. A.C. postdoctoral grant is funded by the ERC Starting Grant 3DICE (grant agreement 336474). M.N.D. acknowledges the financial support of the Center for Space and Habitability (CSH) Fellowship and the IAU Gruber Foundation Fellowship.

%%%%%%%%%%%%%%%%%%%%%%%%%%%%%%%%%%%%%%%%%%%%%%%%%%



\bibliographystyle{aa}
\bibliography{amines} % if your bibtex file is called example.bib

\begin{thebibliography}{59}
\expandafter\ifx\csname natexlab\endcsname\relax\def\natexlab#1{#1}\fi

\bibitem[{{Altwegg} {et~al.}(2016){Altwegg}, {Balsiger}, {Bar-Nun},
  {Berthelier}, {Bieler}, {Bochsler}, {Briois}, {Calmonte}, {Combi}, {Cottin},
  {De Keyser}, {Dhooghe}, {Fiethe}, {Fuselier}, {Gasc}, {Gombosi}, {Hansen},
  {Haessig}, {Ja ckel}, {Kopp}, {Korth}, {Le Roy}, {Mall}, {Marty}, {Mousis},
  {Owen}, {Reme}, {Rubin}, {Semon}, {Tzou}, {Waite}, \& {Wurz}}]{Altwegg2016}
{Altwegg}, K., {Balsiger}, H., {Bar-Nun}, A., {et~al.} 2016, Science Advances,
  2, e1600285

\bibitem[{{Altwegg} {et~al.}(2017){Altwegg}, {Balsiger}, {Berthelier},
  {Bieler}, {Calmonte}, {Fuselier}, {Goesmann}, {Gasc}, {Gombosi}, {Le Roy},
  {de Keyser}, {Morse}, {Rubin}, {Schuhmann}, {Taylor}, {Tzou}, \&
  {Wright}}]{Altwegg2017}
{Altwegg}, K., {Balsiger}, H., {Berthelier}, J.~J., {et~al.} 2017, \mnras, 469,
  S130

\bibitem[{Barrientos {et~al.}(2012)Barrientos, Redondo, Largo, Ray{\'{o}}n, \&
  Largo}]{Barrientos2012}
Barrientos, C., Redondo, P., Largo, L., Ray{\'{o}}n, V.~M., \& Largo, A. 2012,
  \apj, 748, 99

\bibitem[{{Belloche} {et~al.}(2013){Belloche}, {M{\"u}ller}, {Menten},
  {Schilke}, \& {Comito}}]{Belloche2013}
{Belloche}, A., {M{\"u}ller}, H.~S.~P., {Menten}, K.~M., {Schilke}, P., \&
  {Comito}, C. 2013, \aap, 559, A47

\bibitem[{{Blagojevic} {et~al.}(2003){Blagojevic}, {Petrie}, \&
  {Bohme}}]{Blagojevic2003}
{Blagojevic}, V., {Petrie}, S., \& {Bohme}, D.~K. 2003, \mnras, 339, L7

\bibitem[{Bossa {et~al.}(2009)Bossa, Duvernay, Theul{\'{e}}, Borget,
  D'Hendecourt, \& Chiavassa}]{Bossa2009}
Bossa, J.-B., Duvernay, F., Theul{\'{e}}, P., {et~al.} 2009, \aap, 506, 601

\bibitem[{{Caux} {et~al.}(2011){Caux}, {Kahane}, {Castets}, {Coutens},
  {Ceccarelli}, {Bacmann}, {Bisschop}, {Bottinelli}, {Comito}, {Helmich},
  {Lefloch}, {Parise}, {Schilke}, {Tielens}, {van Dishoeck}, {Vastel},
  {Wakelam}, \& {Walters}}]{Caux2011}
{Caux}, E., {Kahane}, C., {Castets}, A., {et~al.} 2011, \aap, 532, A23

\bibitem[{{Codella} {et~al.}(2018){Codella}, {Viti}, {Lefloch}, {Holdship},
  {Bachiller}, {Bianchi}, {Ceccarelli}, {Favre}, {Jim{\'e}nez-Serra}, {Podio},
  \& {Tafalla}}]{Codella2017}
{Codella}, C., {Viti}, S., {Lefloch}, B., {et~al.} 2018, \mnras, 474, 5694

\bibitem[{Congiu {et~al.}(2012)Congiu, Fedoseev, Ioppolo, Dulieu, Chaabouni,
  Baouche, Lemaire, Laffon, Parent, Lamberts, Cuppen, \&
  Linnartz}]{Congiu2012a}
Congiu, E., Fedoseev, G., Ioppolo, S., {et~al.} 2012, \apj, 750, L12

\bibitem[{Coutens {et~al.}(2016)Coutens, J{\o}rgensen, van~der Wiel,
  M{\"{u}}ller, Lykke, Bjerkeli, Bourke, Calcutt, Drozdovskaya, Favre, Fayolle,
  Garrod, Jacobsen, Ligterink, {\"{O}}berg, Persson, van Dishoeck, \&
  Wampfler}]{Coutens2016}
Coutens, A., J{\o}rgensen, J.~K., van~der Wiel, M. H.~D., {et~al.} 2016,
  A{\&}A, 590, L6

\bibitem[{{Crockett} {et~al.}(2014){Crockett}, {Bergin}, {Neill}, {Favre},
  {Schilke}, {Lis}, {Bell}, {Blake}, {Cernicharo}, {Emprechtinger},
  {Esplugues}, {Gupta}, {Kleshcheva}, {Lord}, {Marcelino}, {McGuire},
  {Pearson}, {Phillips}, {Plume}, {van der Tak}, {Tercero}, \&
  {Yu}}]{Crockett2014}
{Crockett}, N.~R., {Bergin}, E.~A., {Neill}, J.~L., {et~al.} 2014, \apj, 787,
  112

\bibitem[{{Danger} {et~al.}(2011){Danger}, {Borget}, {Chomat}, {Duvernay},
  {Theul{\'e}}, {Guillemin}, {Le Sergeant D'Hendecourt}, \&
  {Chiavassa}}]{Danger2011}
{Danger}, G., {Borget}, F., {Chomat}, M., {et~al.} 2011, \aap, 535, A47

\bibitem[{{Dickens} {et~al.}(1997){Dickens}, {Irvine}, {DeVries}, \&
  {Ohishi}}]{Dickens1997}
{Dickens}, J.~E., {Irvine}, W.~M., {DeVries}, C.~H., \& {Ohishi}, M. 1997,
  \apj, 479, 307

\bibitem[{{Elsila} {et~al.}(2009){Elsila}, {Glavin}, \& {Dworkin}}]{Elsila2009}
{Elsila}, J.~E., {Glavin}, D.~P., \& {Dworkin}, J.~P. 2009, Meteoritics and
  Planetary Science, 44, 1323

\bibitem[{{Fedoseev} {et~al.}(2016){Fedoseev}, {Chuang}, {van Dishoeck},
  {Ioppolo}, \& {Linnartz}}]{Fedoseev2016}
{Fedoseev}, G., {Chuang}, K.-J., {van Dishoeck}, E.~F., {Ioppolo}, S., \&
  {Linnartz}, H. 2016, \mnras, 460, 4297

\bibitem[{{Fedoseev} {et~al.}(2012){Fedoseev}, {Ioppolo}, {Lamberts}, {Zhen},
  {Cuppen}, \& {Linnartz}}]{Fedoseev2012}
{Fedoseev}, G., {Ioppolo}, S., {Lamberts}, T., {et~al.} 2012, \jcp, 137, 054714

\bibitem[{{F{\"o}rstel} {et~al.}(2017){F{\"o}rstel}, {Bergantini},
  {Maksyutenko}, {G{\'o}bi}, \& {Kaiser}}]{Forstel2017}
{F{\"o}rstel}, M., {Bergantini}, A., {Maksyutenko}, P., {G{\'o}bi}, S., \&
  {Kaiser}, R.~I. 2017, \apj, 845, 83

\bibitem[{{Garrod}(2013)}]{Garrod2013}
{Garrod}, R.~T. 2013, \apj, 765, 60

\bibitem[{{Garrod} {et~al.}(2008){Garrod}, {Widicus Weaver}, \&
  {Herbst}}]{Garrod2008}
{Garrod}, R.~T., {Widicus Weaver}, S.~L., \& {Herbst}, E. 2008, \apj, 682, 283

\bibitem[{Goesmann {et~al.}(2015)Goesmann, Rosenbauer, Bredehoft, Cabane,
  Ehrenfreund, Gautier, Giri, Kruger, {Le Roy}, MacDermott, McKenna-Lawlor,
  Meierhenrich, Caro, Raulin, Roll, Steele, Steininger, Sternberg, Szopa,
  Thiemann, \& Ulamec}]{Goesmann2015}
Goesmann, F., Rosenbauer, H., Bredehoft, J.~H., {et~al.} 2015, Science, 349

\bibitem[{{Halfen} {et~al.}(2001){Halfen}, {Apponi}, \& {Ziurys}}]{Halfen2001}
{Halfen}, D.~T., {Apponi}, A.~J., \& {Ziurys}, L.~M. 2001, \apj, 561, 244

\bibitem[{Halfen {et~al.}(2013)Halfen, Ilyushin, \& Ziurys}]{Halfen2013}
Halfen, D.~T., Ilyushin, V.~V., \& Ziurys, L.~M. 2013, \apj, 767, 66

\bibitem[{He {et~al.}(2015)He, Vidali, Lemaire, \& Garrod}]{He2015}
He, J., Vidali, G., Lemaire, J.-L., \& Garrod, R.~T. 2015, \apj, 799, 49

\bibitem[{Holtom {et~al.}(2005)Holtom, Bennett, Osamura, Mason, \&
  Kaiser}]{Holtom2005}
Holtom, P.~D., Bennett, C.~J., Osamura, Y., Mason, N.~J., \& Kaiser, R.~I.
  2005, \apj, 626, 940

\bibitem[{{Ilyushin} {et~al.}(2005){Ilyushin}, {Alekseev}, {Dyubko},
  {Motiyenko}, \& {Hougen}}]{Ilyusin2005}
{Ilyushin}, V.~V., {Alekseev}, E.~A., {Dyubko}, S.~F., {Motiyenko}, R.~A., \&
  {Hougen}, J.~T. 2005, Journal of Molecular Spectroscopy, 229, 170

\bibitem[{{Jonusas} \& {Krim}(2016)}]{Jonusas2016}
{Jonusas}, M. \& {Krim}, L. 2016, \mnras, 459, 1977

\bibitem[{{J{\o}rgensen} {et~al.}(under rev.){J{\o}rgensen}, {Calcutt}, \&
  {M{\"u}ller}}]{Jorgensen2017}
{J{\o}rgensen}, J.~K., {Calcutt}, H., \& {M{\"u}ller}, H.~S.~P. under rev.,
  \aap

\bibitem[{J{\o}rgensen {et~al.}(2016)J{\o}rgensen, van~der Wiel, Coutens,
  Lykke, M{\"{u}}ller, van Dishoeck, Calcutt, Bjerkeli, Bourke, Drozdovskaya,
  Favre, Fayolle, Garrod, Jacobsen, {\"{O}}berg, Persson, \&
  Wampfler}]{jorgensen2016}
J{\o}rgensen, J.~K., van~der Wiel, M. H.~D., Coutens, A., {et~al.} 2016,
  A{\&}A, 595, A117

\bibitem[{Kaifu {et~al.}(1974)Kaifu, Morimoto, Nagane, Akabane, Iguchi, \&
  Takagi}]{Kaifu1974}
Kaifu, N., Morimoto, M., Nagane, K., {et~al.} 1974, \apj, 191, L135

\bibitem[{Kim \& Kaiser(2011)}]{Kim2011}
Kim, Y.~S. \& Kaiser, R.~I. 2011, \apj, 729, 68

\bibitem[{{Kirchhoff} {et~al.}(1973){Kirchhoff}, {Johnson}, \&
  {Lovas}}]{Kirchhoff1973}
{Kirchhoff}, W.~H., {Johnson}, D.~R., \& {Lovas}, F.~J. 1973, Journal of
  Physical and Chemical Reference Data, 2, 1

\bibitem[{{Ligterink} {et~al.}(2017){Ligterink}, {Coutens}, {Kofman},
  {M{\"u}ller}, {Garrod}, {Calcutt}, {Wampfler}, {J{\o}rgensen}, {Linnartz}, \&
  {van Dishoeck}}]{Ligterink2017}
{Ligterink}, N.~F.~W., {Coutens}, A., {Kofman}, V., {et~al.} 2017, \mnras, 469,
  2219

\bibitem[{Ligterink {et~al.}(2015)Ligterink, Tenenbaum, \& van
  Dishoeck}]{Ligterink2015}
Ligterink, N. F.~W., Tenenbaum, E.~D., \& van Dishoeck, E.~F. 2015, \aap, 576,
  A35

\bibitem[{{Liszt} \& {Turner}(1978)}]{Liszt1978}
{Liszt}, H.~S. \& {Turner}, B.~E. 1978, \apjl, 224, L73

\bibitem[{{Lykke} {et~al.}(2017){Lykke}, {Coutens}, {J{\o}rgensen}, {van der
  Wiel}, {Garrod}, {M{\"u}ller}, {Bjerkeli}, {Bourke}, {Calcutt},
  {Drozdovskaya}, {Favre}, {Fayolle}, {Jacobsen}, {{\"O}berg}, {Persson}, {van
  Dishoeck}, \& {Wampfler}}]{Lykke2017}
{Lykke}, J.~M., {Coutens}, A., {J{\o}rgensen}, J.~K., {et~al.} 2017, \aap, 597,
  A53

\bibitem[{{McGonagle} {et~al.}(1990){McGonagle}, {Irvine}, {Minh}, \&
  {Ziurys}}]{Mcgonagle1990}
{McGonagle}, D., {Irvine}, W.~M., {Minh}, Y.~C., \& {Ziurys}, L.~M. 1990, \apj,
  359, 121

\bibitem[{McGuire {et~al.}(2015)McGuire, Carroll, Dollhopf, Crockett, Corby,
  Loomis, {M. Burkhardt}, Shingledecker, Blake, \& Remijan}]{Mcguire2015}
McGuire, B.~A., Carroll, P.~B., Dollhopf, N.~M., {et~al.} 2015, \apj, 812, 76

\bibitem[{{Milam} {et~al.}(2005){Milam}, {Savage}, {Brewster}, {Ziurys}, \&
  {Wyckoff}}]{Milam2005}
{Milam}, S.~N., {Savage}, C., {Brewster}, M.~A., {Ziurys}, L.~M., \& {Wyckoff},
  S. 2005, \apj, 634, 1126

\bibitem[{Morino {et~al.}(2000)Morino, Yamada, Klein, Belov, Winnewisser,
  Bocquet, Wlodarczak, Lodyga, \& Kreglewski}]{Morino2000}
Morino, I., Yamada, K., Klein, H., {et~al.} 2000, Journal of Molecular
  Structure, 517, 367

\bibitem[{{M{\"u}ller} {et~al.}(2005){M{\"u}ller}, {Schl{\"o}der}, {Stutzki},
  \& {Winnewisser}}]{muller2005}
{M{\"u}ller}, H.~S.~P., {Schl{\"o}der}, F., {Stutzki}, J., \& {Winnewisser}, G.
  2005, Journal of Molecular Structure, 742, 215

\bibitem[{{M{\"u}ller} {et~al.}(2001){M{\"u}ller}, {Thorwirth}, {Roth}, \&
  {Winnewisser}}]{muller2001}
{M{\"u}ller}, H.~S.~P., {Thorwirth}, S., {Roth}, D.~A., \& {Winnewisser}, G.
  2001, \aap, 370, L49

\bibitem[{Neill {et~al.}(2014)Neill, Bergin, Lis, Schilke, Crockett, Favre,
  Emprechtinger, Comito, Qin, Anderson, Burkhardt, Chen, Harris, Lord, McGuire,
  McNeill, Monje, Phillips, Steber, Vasyunina, \& Yu}]{Neill2014}
Neill, J.~L., Bergin, E.~A., Lis, D.~C., {et~al.} 2014, \apj, 789, 8

\bibitem[{Nummelin {et~al.}(2000)Nummelin, Bergman, Hjalmarson, Friberg,
  Irvine, Millar, Ohishi, \& Saito}]{Nummelin2000}
Nummelin, A., Bergman, P., Hjalmarson, A., {et~al.} 2000, Doktorsavhandlingar
  vid Chalmers Tekniska Hogskola, 2

\bibitem[{{Pagani} {et~al.}(2017){Pagani}, {Favre}, {Goldsmith}, {Bergin},
  {Snell}, \& {Melnick}}]{Pagani2017}
{Pagani}, L., {Favre}, C., {Goldsmith}, P.~F., {et~al.} 2017, \aap, 604, A32

\bibitem[{{Persson} {et~al.}(2018){Persson}, {J{\o}rgensen}, {M{\"u}ller},
  {Coutens}, {van Dishoeck}, {Taquet}, {Calcutt}, {van der Wiel}, {Bourke}, \&
  {Wampfler}}]{Persson2017}
{Persson}, M.~V., {J{\o}rgensen}, J.~K., {M{\"u}ller}, H.~S.~P., {et~al.} 2018,
  \aap, 610, A54

\bibitem[{{Pickett} {et~al.}(1979){Pickett}, {Cohen}, {Waters}, \&
  {Phillips}}]{Pickett1979}
{Pickett}, H.~M., {Cohen}, E.~A., {Waters}, J.~W., \& {Phillips}, T.~G. 1979,
  34th International Symposium on Molecular Spectroscopy, Columbus, OH, USA.

\bibitem[{{Pickett} {et~al.}(1998){Pickett}, {Poynter}, {Cohen}, {Delitsky},
  {Pearson}, \& {M{\"u}ller}}]{Pickett1998}
{Pickett}, H.~M., {Poynter}, R.~L., {Cohen}, E.~A., {et~al.} 1998, \jqsrt, 60,
  883

\bibitem[{Pulliam {et~al.}(2012)Pulliam, McGuire, \& Remijan}]{Pulliam2012}
Pulliam, R.~L., McGuire, B.~A., \& Remijan, A.~J. 2012, \apj, 751, 1

\bibitem[{{Snow} {et~al.}(2007){Snow}, {Orlova}, {Blagojevic}, \&
  {Bohme}}]{Snow2007}
{Snow}, J.~L., {Orlova}, G., {Blagojevic}, V., \& {Bohme}, D.~K. 2007, J. Am.
  Chem. Soc., 129, 9910–9917

\bibitem[{{Suzuki} {et~al.}(2016){Suzuki}, {Ohishi}, {Hirota}, {Saito},
  {Majumdar}, \& {Wakelam}}]{Suzuki2016}
{Suzuki}, T., {Ohishi}, M., {Hirota}, T., {et~al.} 2016, \apj, 825, 79

\bibitem[{Theule {et~al.}(2011)Theule, Borget, Mispelaer, Danger, Duvernay,
  Guillemin, \& Chiavassa}]{Theule2011}
Theule, P., Borget, F., Mispelaer, F., {et~al.} 2011, \aap, 534, A64

\bibitem[{{Ting} {et~al.}(2014){Ting}, {Chang}, {Chen}, {Chen}, {Shy},
  {Drouin}, \& {Daly}}]{Ting2014}
{Ting}, W.-J., {Chang}, C.-H., {Chen}, S.-E., {et~al.} 2014, Journal of the
  Optical Society of America B Optical Physics, 31, 1954

\bibitem[{{Turner}(1991)}]{Turner1991}
{Turner}, B.~E. 1991, \apjs, 76, 617

\bibitem[{{Visser} {et~al.}(2011){Visser}, {Doty}, \& {van
  Dishoeck}}]{Visser2011}
{Visser}, R., {Doty}, S.~D., \& {van Dishoeck}, E.~F. 2011, \aap, 534, A132

\bibitem[{{Wilson}(1999)}]{Wilson1999}
{Wilson}, T.~L. 1999, Reports on Progress in Physics, 62, 143

\bibitem[{{Woon}(2002)}]{Woon2002}
{Woon}, D.~E. 2002, \apjl, 571, L177

\bibitem[{{Y{\i}ld{\i}z} {et~al.}(2013){Y{\i}ld{\i}z}, {Acharyya}, {Goldsmith},
  {van Dishoeck}, {Melnick}, {Snell}, {Liseau}, {Chen}, {Pagani}, {Bergin},
  {Caselli}, {Herbst}, {Kristensen}, {Visser}, {Lis}, \& {Gerin}}]{Yildiz2013}
{Y{\i}ld{\i}z}, U.~A., {Acharyya}, K., {Goldsmith}, P.~F., {et~al.} 2013, \aap,
  558, A58

\bibitem[{{Zheng} \& {Kaiser}(2010)}]{Zheng2010}
{Zheng}, W. \& {Kaiser}, R.~I. 2010, Journal of Physical Chemistry A, 114, 5251

\bibitem[{{Ziurys} {et~al.}(1994){Ziurys}, {Apponi}, {Hollis}, \&
  {Snyder}}]{Ziurys1994}
{Ziurys}, L.~M., {Apponi}, A.~J., {Hollis}, J.~M., \& {Snyder}, L.~E. 1994,
  \apjl, 436, L181

\end{thebibliography}

%%%%%%%%%%%%%%%%%%%%%%%%%%%%%%%%%%%%%%%%%%%%%%%%%%

%%%%%%%%%%%%%%%%% APPENDICES %%%%%%%%%%%%%%%%%%%%%

\newpage
\appendix

\section{Spectroscopic data}
\label{ap:spec}

Table \ref{tab.lines} lists the transitions of NO, N$_{2}$O and CH$_{2}$NH detected towards IRAS 16293B. Columns seven through ten list the line FWHM, peak velocity $V_\mathrm{peak}$, line opacity $\tau$ and peak brightness temperature $T_{\rm peak}$ for each transition, based on the synthetic spectra at 100~K. The final column indicates if a line is partially blended with a neighboring line. The error on $V_\mathrm{peak}$ is the velocity resolution of 0.2~km~s$^{-1}$. The line FWHM is kept as a fixed parameter at 1~km~s$^{-1}$. The error in the peak brightness is given by the calibration uncertainty and RMS noise, as indicated in Section \ref{sec.obs}.

\begin{sidewaystable*}[!t]
\center
\caption{Detected lines of NO, N$_{2}$O and CH$_{2}$NH. \label{tab.lines}}
         \begin{tabular}{l c c c c c c c c c c}
            \hline
            \hline
            
            Molecules & Transitions & Frequency & $E_\mathrm{up}$ & $A_\mathrm{ij}$ & $g_\mathrm{up}$ & FWHM$^{a}$ & $V_\mathrm{peak}$$^{a}$ & $\tau^{a}$ & $T_{\rm peak}^{a}$ & Blended$^{b}$ \\
            & & (MHz) & (K) & (s$^{-1}$) &  & (km s$^{-1}$) & (km s$^{-1}$) & & (K) & \\
            
            \hline
            
            NO & 4 $-$1 7/2 9/2 -- 3 \phantom{$-$}1 5/2 7/2 & 350 689.49 & 36 & 5.42$\times$10$^{-6}$ & 10 & 1.0 & 2.5 $\pm$0.2 & 1.1$\times$10$^{-1}$ & 4.0 $\pm$0.6& B \\
            NO & 4 $-$1 7/2 7/2 -- 3 \phantom{$-$}1 5/2 5/2 & 350 690.76 & 36 & 4.97$\times$10$^{-6}$ & 8 & 1.0 & 2.5 $\pm$0.2 & 7.8$\times$10$^{-2}$ & 3.0 $\pm$0.5 & B \\
            NO & 4 $-$1 7/2 5/2 -- 3 \phantom{$-$}1 5/2 3/2 & 350 694.77 & 36 & 4.81$\times$10$^{-6}$ & 6 & 1.0 & 2.5 $\pm$0.2 & 5.7$\times$10$^{-2}$ & 2.2 $\pm$0.5 & U \\
            NO & 4\phantom{$-$} 1 7/2 9/2 -- 3 $-$1 5/2 7/2 & 351 043.52 & 36 & 5.43$\times$10$^{-6}$ & 10 & 1.0 & 2.5 $\pm$0.2 & 1.1$\times$10$^{-1}$ & 4.0 $\pm$0.6 & B \\
            NO & 4 \phantom{$-$}1 7/2 7/2 -- 3 $-$1 5/2 5/2 & 351 051.46 & 36 & 4.99$\times$10$^{-6}$ & 8 & 1.0 & 2.5 $\pm$0.2 & 7.8$\times$10$^{-2}$ & 3.0 $\pm$0.5 & U \\
            NO & 4 \phantom{$-$}1 7/2 5/2 -- 3 $-$1 5/2 3/2 & 351 051.70 & 36 & 4.83$\times$10$^{-6}$ & 6 & 1.0 & 2.5 $\pm$0.2 & 5.7$\times$10$^{-2}$ & 2.2 $\pm$0.5 & U \\
            
            \hline
            N$_{2}$O & 14\,--\,13 & 351 667.81 & 127 & 6.32$\times$10$^{-6}$ & 29 & 1.0 & 2.5 $\pm$0.2 & 5.3$\times$10$^{-1}$ & 16.3 $\pm$1.6 & U \\
            \hline
            
            CH$_{2}$NH & 5 1 4 5 -- 4 1 3 4 & 332 573.04 & 56 & 3.15$\times$10$^{-4}$ & 11 & 1.0 & 2.7 $\pm$0.2 & 6.3$\times$10$^{-2}$ & 2.3 $\pm$0.5 & B \\
            CH$_{2}$NH & 5 1 4 6 -- 4 1 3 5 & 332 573.07 & 56 & 3.28$\times$10$^{-4}$ & 13 & 1.0 & 2.7 $\pm$0.2 & 7.8$\times$10$^{-2}$ & 2.8 $\pm$0.5 & B \\
            CH$_{2}$NH & 5 1 4 4 -- 4 1 3 3 & 332 573.11 & 56 & 3.12$\times$10$^{-4}$ & 9 & 1.0 & 2.7 $\pm$0.2 & 5.1$\times$10$^{-2}$ & 1.9 $\pm$0.5 & B \\
            CH$_{2}$NH & 3 1 3 3 -- 2 0 2 3 & 340 353.05 & 26 & 3.44$\times$10$^{-5}$ & 7 & 1.0 & 2.7 $\pm$0.2 & 5.7$\times$10$^{-3}$ & 0.2 $\pm$0.3 & U \\
            CH$_{2}$NH & 3 1 3 3 -- 2 0 2 2 & 340 353.37 & 26 & 2.75$\times$10$^{-4}$ & 7 & 1.0 & 2.7 $\pm$0.2 & 4.6$\times$10$^{-2}$ & 1.7 $\pm$0.4 & U \\
            CH$_{2}$NH & 3 1 3 4 -- 2 0 2 3 & 340 354.31 & 26 & 3.09$\times$10$^{-4}$ & 9 & 1.0 & 2.7 $\pm$0.2 & 6.6$\times$10$^{-2}$ & 2.5 $\pm$0.4 & U \\
            CH$_{2}$NH & 3 1 3 2 -- 2 0 2 1 & 340 354.58 & 26 & 2.60$\times$10$^{-4}$ & 5 & 1.0 & 2.7 $\pm$0.2 & 3.1$\times$10$^{-2}$ & 1.2 $\pm$0.3 & U \\
            CH$_{2}$NH & 3 1 3 2 -- 2 0 2 2 & 340 355.08 & 26 & 4.81$\times$10$^{-5}$ & 5 & 1.0 & 2.7 $\pm$0.2 & 5.7$\times$10$^{-3}$ & 0.2 $\pm$0.3 & U \\
            CH$_{2}$NH & \phantom{2}10 1 9 9 -- 10 0 10 9 & 351 453.85 & 183 & 3.61$\times$10$^{-4}$ & 19 & 1.0 & 2.7 $\pm$0.2 & 3.2$\times$10$^{-2}$ & 1.2 $\pm$0.5 & B \\
            CH$_{2}$NH & \phantom{2}10 1 9 11 -- 10 0 10 11 & 351 454.02 & 183 & 3.62$\times$10$^{-4}$ & 23 & 1.0 & 2.7 $\pm$0.2 & 3.8$\times$10$^{-2}$ & 1.5 $\pm$0.5 & B \\
            CH$_{2}$NH & 10 1 9 10 -- 10 0 10 9 & 351 454.43 & 183 & 3.30$\times$10$^{-6}$ & 21 & 1.0 & 2.7 $\pm$0.2 & 3.2$\times$10$^{-4}$ & 0.01 $\pm$0.4 & B \\
            CH$_{2}$NH & \phantom{2}10 1 9 10 -- 10 0 10 11 & 351 454.55 & 183 & 3.30$\times$10$^{-6}$ & 21 & 1.0 & 2.7 $\pm$0.2 & 3.2$\times$10$^{-4}$ & 0.01 $\pm$0.4 & B \\
            CH$_{2}$NH & \phantom{2}10 1 9 11 -- 10 0 10 11 & 351 454.02 & 183 & 3.62$\times$10$^{-4}$ & 23 & 1.0 & 2.7 $\pm$0.2 & 3.2$\times$10$^{-4}$ & 0.01 $\pm$0.4 & B \\
            CH$_{2}$NH & \phantom{2}10 1 9 11 -- 10 0 10 10 & 351 455.17 & 183 & 3.02$\times$10$^{-6}$ & 23 & 1.0 & 2.7 $\pm$0.2 & 3.2$\times$10$^{-4}$ & 0.01 $\pm$0.4 & B \\
            CH$_{2}$NH & \phantom{2}10 1 9 10 -- 10 0 10 10 & 351 455.70 & 183 & 3.58$\times$10$^{-4}$ & 21 & 1.0 & 2.7 $\pm$0.2 & 3.5$\times$10$^{-2}$ & 1.4 $\pm$0.5 & B \\
            CH$_{2}$NH & 6 3 4 6 -- 7 2 5 7 & 360 293.85 & 135 & 6.91$\times$10$^{-5}$ & 13 & 1.0 & 2.7 $\pm$0.2 & 6.3$\times$10$^{-3}$& 0.3 $\pm$0.3 & B \\
            CH$_{2}$NH & 6 3 4 7 -- 7 2 5 8 & 360 294.05 & 135 & 6.93$\times$10$^{-5}$ & 15 & 1.0 & 2.7 $\pm$0.2 & 7.3$\times$10$^{-3}$ & 0.3 $\pm$0.3 & B \\
            CH$_{2}$NH & 6 3 4 5 -- 7 2 5 6 & 360 294.08 & 135 & 7.05$\times$10$^{-5}$ & 11 & 1.0 & 2.7 $\pm$0.2 & 5.5$\times$10$^{-3}$ & 0.2 $\pm$0.3 & B \\
            
            \hline
            
         \end{tabular}     
          
         \tablefoot{Quantum numbers are given as N$'$\,$P\Lambda'$\,J$'$\,F$'$ -- N$''$\,$P\Lambda''$\,J$''$\,F$''$ for NO, J$'$ -- J$''$ for N$_2$O and J$'$\,K$_a'$\,K$_c'$\,F$'$ -- J$''$\,K$_a''$\,K$_c''$\,F$''$ for CH$_2$NH. $^{a}$line width, peak line velocity, opacity and peak brightness temperature are taken from the synthetic fit of each molecule at $T_{\rm ex}$ = 100~K. The spectral resolution of the data is 0.2 km s$^{-1}$. $^{b}$B = Blended, U = Unblended. Column covers blends with other species, not blends with transitions of the same species.}
\end{sidewaystable*}

\section{Additional fit spectra of CH$_{2}$NH and NO}
\label{ap:add_spec}

\begin{figure*}[!t] 
 \begin{center} 
 \includegraphics[width=\hsize]{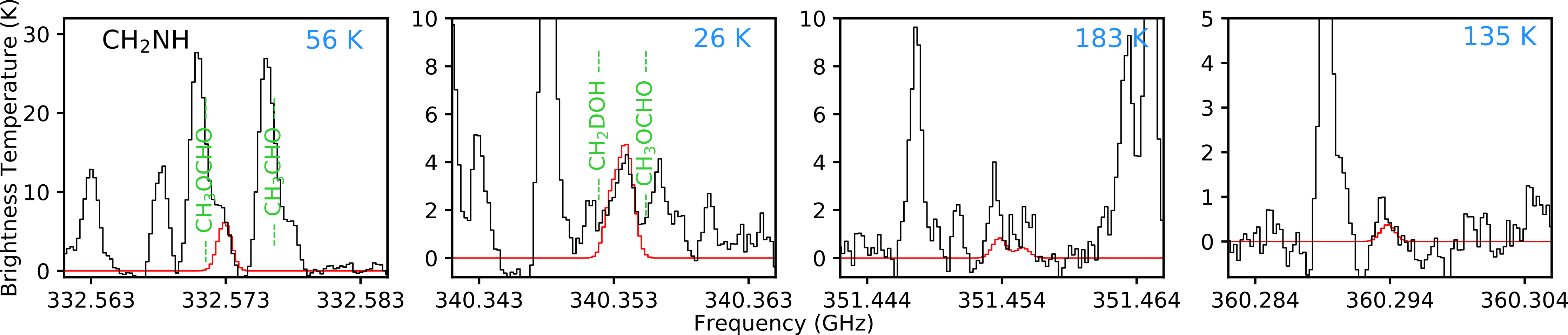}
 \includegraphics[width=\hsize]{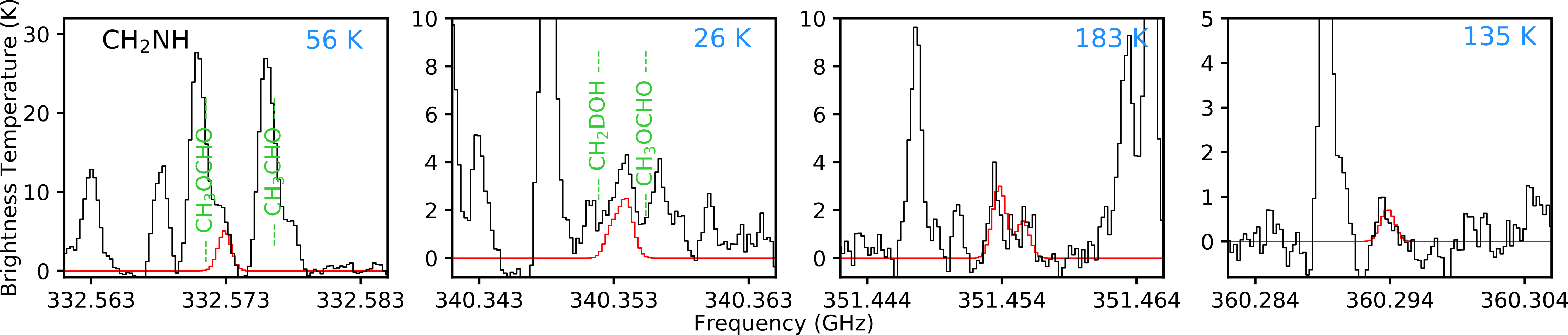}
 \end{center} 
 \caption{Fit models (red) of CH$_{2}$NH at $T_{\rm ex}$ = 50~K (top) and $T_{\rm ex}$ = 150~K (bottom) overplotted on the PILS data (black) and other detected species in PILS given in green. Upper state energies of the CH$_{2}$NH transitions are indicated in blue.
 \label{fig:Tex_CH2NH}} 
 \end{figure*}

Figure \ref{fig:Tex_CH2NH} shows CH$_{2}$NH synthetic spectra at $T_{\rm ex}$ = 50 and 150~K. Best fit column densities of 7$\times$10$^{14}$ and 8$\times$10$^{14}$ cm$^{-2}$, respectively, are found. However, clear discrepancies in the fits exist, which are especially visible for the transitions at 340.354 and 351.454 GHz.

\begin{figure*}[!t] 
 \begin{center} 
 \includegraphics[width=\hsize]{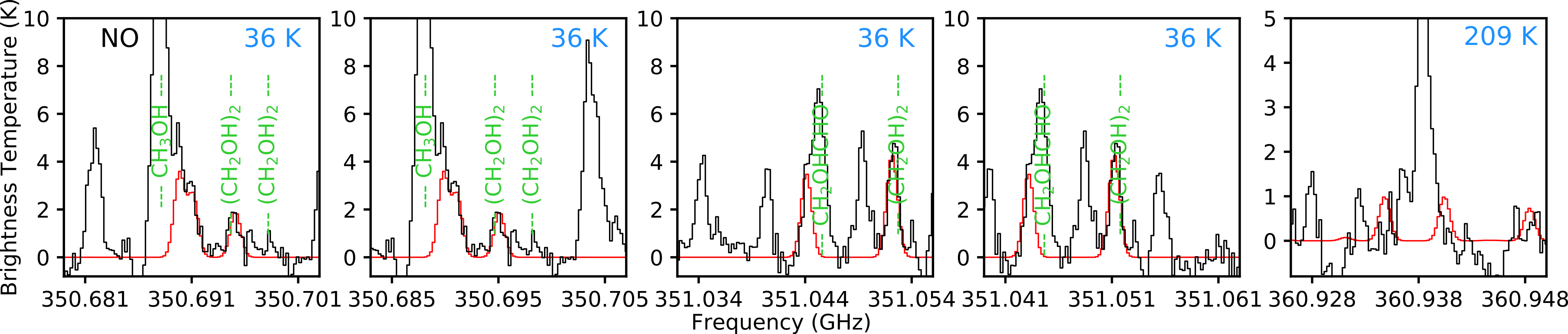}
 \end{center} 
 \caption{Fit models (red) of NO at $T_{\rm ex}$ = 150~K overplotted on the PILS data (black) and other detected species in PILS given in green. Upper state energies of the NO transitions are indicated in blue.
 \label{fig:Tex_NO}} 
 \end{figure*}

Figure~\ref{fig:Tex_NO} shows the NO synthetic spectrum at $T_{\rm ex}$ = 150~K and $N_{\rm tot}$ = 2.3$\times$10$^{16}$ cm$^{-2}$. The transitions at $E_{\rm up}$ = 36~K are well fitted, however, clear anti-coincidences start to arise at the transitions of $E_{\rm up}$ = 209~K, as can be seen for the transitions at 360.935 and 360.941 GHz. Excitation temperatures for NO are therefore constrained to $T_{\rm ex}$ < 150~K.

%%%%%%%%%%%%%%%%%%%%%%%%%%%%%%%%%%%%%%%%%%%%%%%%%%%%%%%%%%%%%%%%%%%%%%%%%%%%%%%%

\section{N$_{2}$O fit of TIMASSS data set}
\label{ap:timasss_n2o}

\begin{figure*}[!t] 
 \begin{center} 
 \includegraphics[width=\hsize]{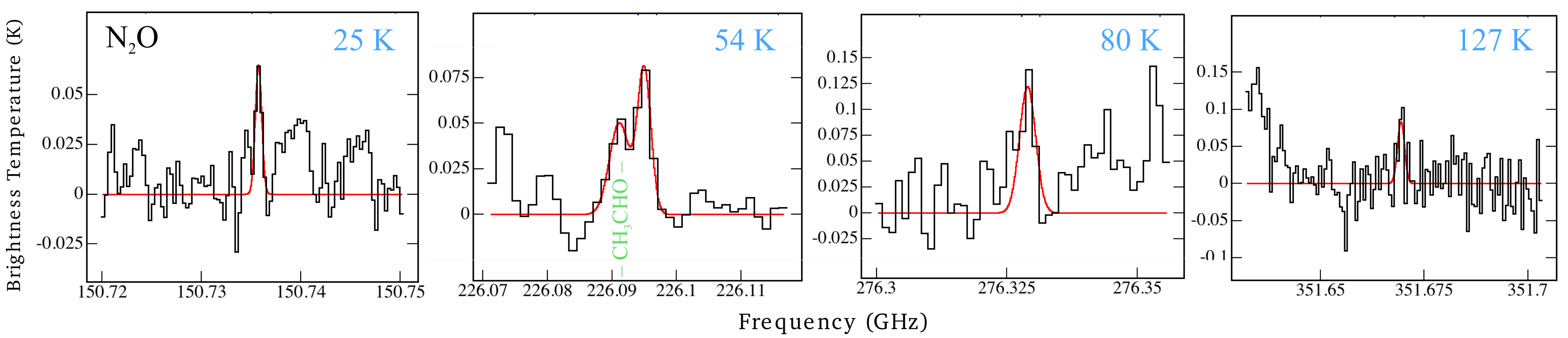}
 \end{center} 
 \caption{Gaussian fits (red) of unblended N$_{2}$O transitions in the single-dish TIMASSS data (black) by \citet{Caux2011}. Upper state energies of the transitions are indicated in blue.
 \label{fig:timasss_n2o}} 
 \end{figure*}

\begin{figure}[!t] 
 \begin{center} 
 \includegraphics[width=\hsize]{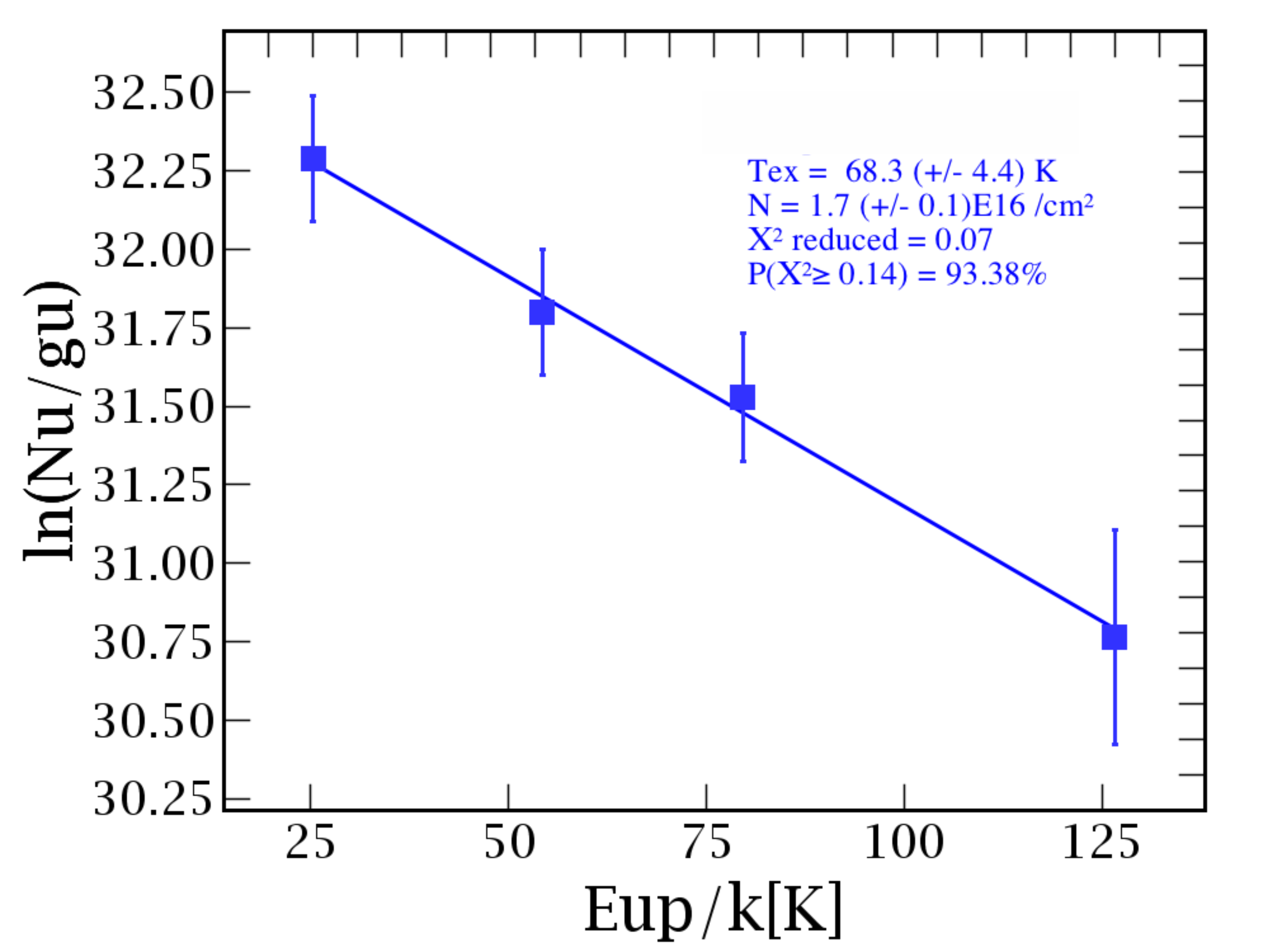}
 \end{center} 
 \caption{Rotational temperature diagram of the four unblended N$_{2}$O transitions found in the TIMASSS dataset, resulting in $T_{\rm ex}$ = 68~K and $N$ = 1.7$\times$10$^{16}$ cm$^{-2}$, for a source size of 1\farcs6.
 \label{fig:rtd_n2o}} 
 \end{figure}

Figure \ref{fig:timasss_n2o} shows the four unblended lines of N$_{2}$O detected in the TIMASSS survey. All these transitions are velocity shifted to $V_{\rm peak}$ = 2.5 km\,s$^{-1}$. Gaussian profiles were fitted and line widths are found between 1.4 and 3.3 km\,s$^{-1}$, consistent with other FWHMs found in TIMASSS data for source B \citep{Caux2011}. The N$_{2}$O transition at 226.95 GHz is fitted with a second gaussian to properly account for the contribution of a nearby methylformate (CH$_{3}$OCHO) transition. From the fits a rotational temperature diagram is created, shown in Fig. \ref{fig:rtd_n2o}. The diagram gives $T_{\rm ex}$ = 68~K and $N$ = 1.7$\times$10$^{16}$ cm$^{-2}$, using a source size of 1\farcs6 (see text). Variations in source size will change the excitation temperature and column density slightly.

%%%%%%%%%%%%%%%%%%%%%%%%%%%%%%%%%%%%%%%%%%%%%%%%%%%%%%%%%%%%%%%%%%%%%%%%%%%%%%%%

\section{Upper limit column densities of NH$_{2}$OH and CH$_{3}$NH$_{2}$}
\label{ap:non}

\begin{figure}[!t]
  \centering
  \includegraphics[width=\hsize]{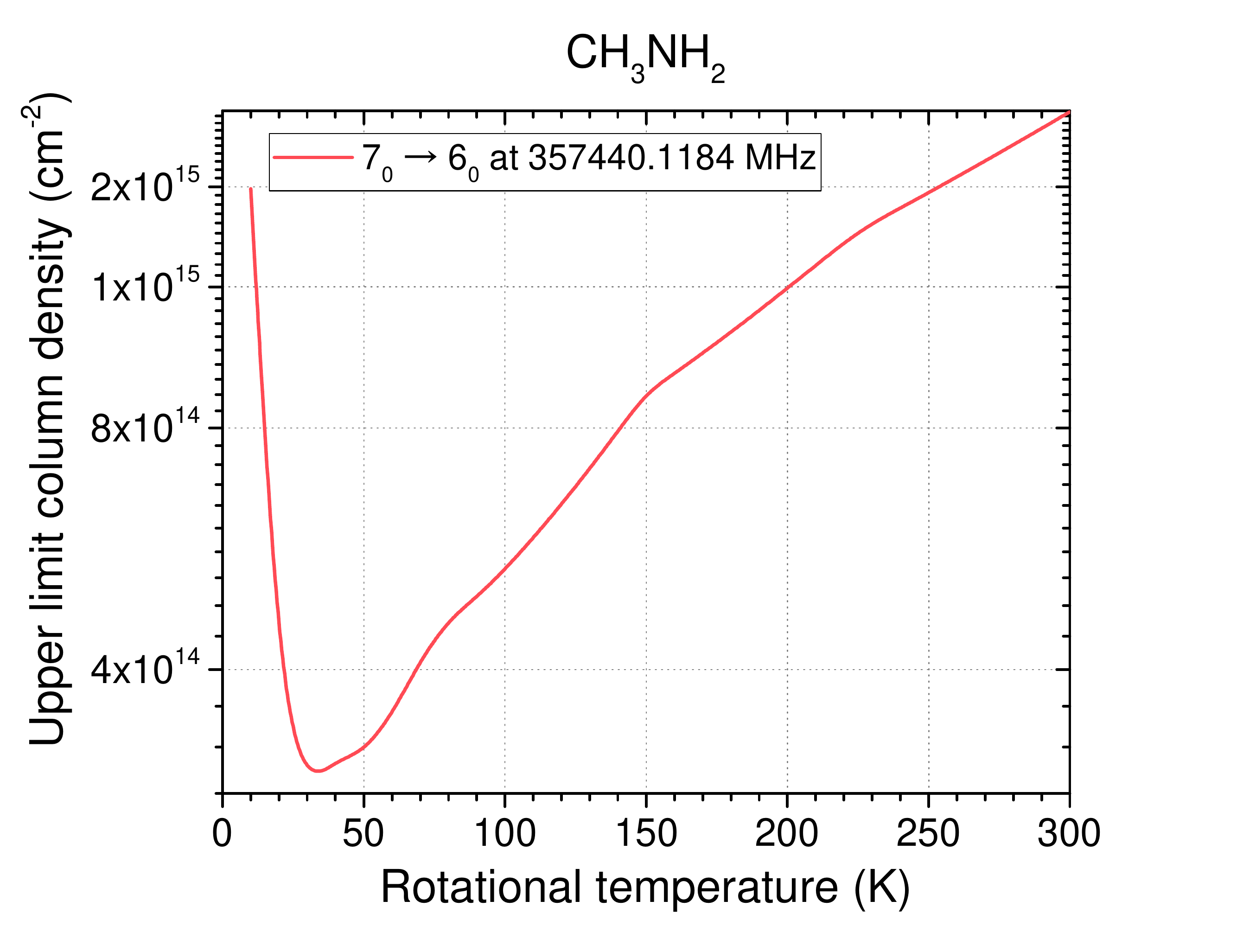}
  \caption{Upper limit column density for the strongest CH$_3$NH$_2$ transition as function of $T_{\rm ex}$. A 3$\sigma$ value of 23\,mJy\,km\,s$^{-1}$ is used.}
  \label{fig:ch3nh2_up}
\end{figure}

\begin{figure}[!t]
\centering
  \includegraphics[width=\columnwidth]{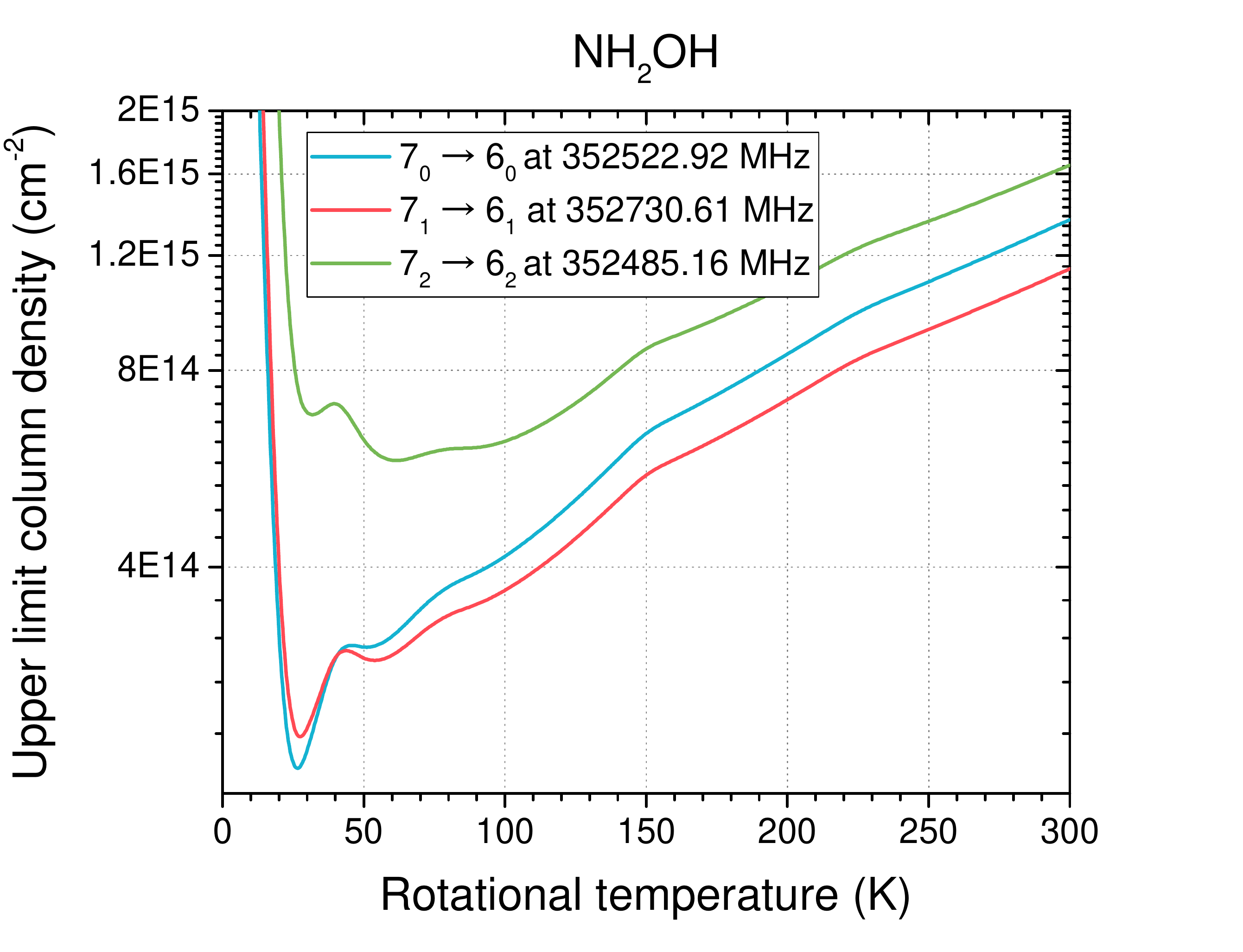}
  \caption{Upper limit column densities for the three strongest NH$_{2}$OH transitions as function of $T_{\rm ex}$. 3$\sigma$ values of 27, 21 and 27\,mJy\,km\,s$^{-1}$ are used for the respective lines.}
  \label{fig:nh2oh_up}
\end{figure}

Upper limit column densities of NH$_{2}$OH and CH$_{3}$NH$_{2}$ have been determined for excitation temperatures between 10 and 300~K. These are plotted in Figure \ref{fig:ch3nh2_up} and \ref{fig:nh2oh_up}. The upper limit column densities of NH$_{2}$OH are determined on the three transitions:  7$_{0}$ $\rightarrow$ 6$_{0}$, 7$_{1}$ $\rightarrow$ 6$_{1}$, 7$_{2}$ $\rightarrow$ 6$_{2}$ transitions at 352 522, 352 730 and 352 485 MHz, respectively. These transitions are modelled with a 3$\sigma$ line intensity of 27, 21 and 27\,mJy\,km\,s$^{-1}$, respectively. The upper limit column densities of CH$_{3}$NH$_{2}$ are determined on the 6$_{5}$ $\rightarrow$ 5$_{0}$ transition at 357 440 MHz for a 3$\sigma$ line intensity of 23\,mJy\,km\,s$^{-1}$.

%\begin{figure}[H]
%\centering
%\begin{minipage}{.45\textwidth}
%  \centering
%  \includegraphics[width=\hsize]{NH2OH_uplim.pdf}
%  \captionof{figure}{Upper limit column densities for the three strongest NH$_{2}$OH transitions. 3$\sigma$ values of 27, 21 and 27\,mJy\,km\,s$^{-1}$ are used for the respective lines.}
%  \label{fig:nh2oh_up}
%\end{minipage}%
%\hspace{1cm}
%\begin{minipage}{.45\textwidth}
%  \centering
%  \includegraphics[width=\hsize]{CH3NH2_uplim.pdf}
%  \captionof{figure}{Upper limit column density for the strongest CH$_3$NH$_2$ transition. A 3$\sigma$ value of 23\,mJy\,km\,s$^{-1}$ is used.}
%  \label{fig:ch3nh2_up}
%\end{minipage}
%\end{figure}

%%%%%%%%%%%%%%%%%%%%%%%%%%%%%%%%%%%%%%%%%%%%%%%%%%

\section{Comparison of NH$_{2}$OH upper limits}

\begin{table*}
     \caption[]{Upper limit abundance ratios for hydroxylamine in IRAS16293--2422.}
         \label{tab.nh2oh}
         $$
         \begin{tabular}{l c c l}
            \hline
            \hline
            
            Source & NH$_{2}$OH/NO & NH$_{2}$OH/H$_{2}$ & Reference \\
            
            \hline
            
          	IRAS16293-2422B	& $\leq$0.025 & $\leq$3.1$\times$10$^{-11}$ & This work	\\
           	\hline
            
			Orion KL &  -- & $\leq$3$\times$10$^{-11}$ & \citet{Pulliam2012} \\
			Orion S & -- & $\leq$9$\times$10$^{-11}$ & \citet{Pulliam2012}\\
			IRC+10216 & -- & $\leq$3$\times$10$^{-9}$ & \citet{Pulliam2012}\\
			Sgr B2(OH) & -- & $\leq$3$\times$10$^{-11}$ & \citet{Pulliam2012}\\
			Sgr B2(N) & -- & $\leq$8$\times$10$^{-12}$ & \citet{Pulliam2012}\\
			W51M & -- & $\leq$4$\times$10$^{-11}$ & \citet{Pulliam2012}\\
			W3IRS5 & -- & $\leq$3$\times$10$^{-11}$ & \citet{Pulliam2012}\\
			L1157B1 & -- & $\leq$1.4$\times$10$^{-8}$ & \citet{Mcguire2015}\\
			L1157B2 & -- & $\leq$1.5$\times$10$^{-8}$ & \citet{Mcguire2015}\\
            \hline
            Dark cloud model & -- & 7$\times$10$^{-9}$ & \citet{Fedoseev2012} \\
            Hot core model & 7.3-110$\times$10$^{-5}$ & 1.6-17$\times$10$^{-10}$ & \citet{Garrod2013}$^{a}$ \\
            
            \hline        
         \end{tabular}     
         $$  
         \tablefoot{$^{a}$Abundance ranges taken from the F(ast), M(edium) and S(low) warm-up models.} 
\end{table*}

%%%%%%%%%%%%%%%%%%%%%%%%%%%%%%%%%%%%%%%%%%%%%%%%%%

% Don't change these lines
%\bsp	% typesetting comment
\label{lastpage}

\end{document}